\documentclass[a4paper,11pt]{article}
\pdfoutput=1 
\usepackage{jheppub,bm,color,bbold} 
\usepackage[T1]{fontenc}

\DeclareMathOperator\tr{tr}
\DeclareMathOperator\diag{diag}
\DeclareMathOperator\ee{\text{e}}
\newcommand{\dd}{\mathrm{d}}

\newcommand{\RR}{\mathbb{R}}
\newcommand{\Q}{Q}

\newcommand{\SU}{\text{SU}}

\newcommand{\U}{\text{U}}
\newcommand{\calO}{\mathcal{O}}

\newcommand{\beq}{\begin{eqnarray}}
\newcommand{\eeq}{\end{eqnarray}}
\newcommand{\der}{\partial}
\def\ba#1\ea{\begin{align}#1\end{align}}
\def\akakko#1{\left\langle #1 \right\rangle}
\def\mkakko#1{\left( #1 \right)}
\def\ckakko#1{\left\{ #1 \right\}}
\def\kkakko#1{\left[ #1 \right]}
\renewcommand{\epsilon}{\varepsilon}
\renewcommand{\bar}[1]{\overline{#1}}

\preprint{RIKEN-QHP-188}

\title{\boldmath 
$\U(1)$ axial symmetry and Dirac spectra in QCD at high temperature}

\author[a]{Takuya Kanazawa}
\author[b]{and Naoki Yamamoto}

\affiliation[a]{iTHES Research Group and Quantum Hadron Physics Laboratory, RIKEN, Wako, Saitama 351-0198, Japan}
\affiliation[b]{Department of Physics, Keio University, Yokohama 223-8522, Japan}

\emailAdd{takuya.kanazawa@riken.jp}
\emailAdd{nyama@rk.phys.keio.ac.jp}

\abstract{We derive some exact results concerning the anomalous $\U(1)_A$ symmetry in 
the chirally symmetric phase of QCD at high temperature. 
We discuss the importance of topology and finite-volume effects on the $\U(1)_A$ symmetry violation 
characterized by the difference of chiral susceptibilities. In particular, we present 
a reliable method to measure the anomaly strength in lattice simulations with fixed topology. 
We also derive new spectral sum rules and a novel Banks--Casher-type relation. 
Through our spectral analysis we arrive at a simple alternative proof of the Aoki--Fukaya--Taniguchi 
``theorem'' on the effective restoration of the $\U(1)_A$ symmetry at high temperature. 
}

\begin{document}
\maketitle
\flushbottom
\section{Introduction}
\label{sc:intro}

The physics of the $\U(1)$ axial symmetry in quantum chromodynamics (QCD) 
has been a subject of intensive research over many years. 
While the QCD Lagrangian is invariant under $\U(N_f)_R \times \U(N_f)_L$ at the classical level, 
the $\U(1)_A$ symmetry is broken by quantum effects 
\cite{Adler:1969gk} (see (\ref{anomaly})). In fact, experimentally observed 
hadron spectra in the vacuum do not fully respect the $\U(N_f)_R \times \U(N_f)_L$ symmetry: 
the $\eta'$ meson concerning the $\U(1)_A$ symmetry is much heavier than 
the other pseudoscalar mesons associated with chiral symmetry breaking. 
This $\U(1)_A$ problem was resolved 
\cite{Hooft:1976up, Hooft:1976fv} through the discovery of 
nonperturbative topological excitations in the Euclidean spacetime, called instantons.

Recently, much attention has been focused on another ``$\U(1)_A$ puzzle'' in 
QCD at $T>T_c$\,, where $T_c$ is the (pseudo)critical temperature of the chiral transition. 
Namely, even though the $\U(1)_A$ anomaly relation \eqref{anomaly} itself does not 
receive any modification at finite temperature \cite{Itoyama:1982up},%
\footnote{We noticed that the tensor decomposition of the anomalous correlation 
function at finite temperature in \cite{Itoyama:1982up} is incomplete, and that 
many more terms need to be included; see Appendix \ref{app:tensor}.
This does not, however, affect the conclusion of \cite{Itoyama:1982up} that
the anomaly relation (\ref{anomaly}) remains unchanged at finite temperature.}
the $\U(1)_A$ symmetry could be \emph{effectively} restored at the level of  
mesonic two-point (or higher-point) correlation functions \cite{Shuryak:1993ee}. 
This problem is of particular interest due to the role played by the axial anomaly in 
determining the order of chiral phase transition \cite{Pisarski:1983ms}. 
It was claimed by Cohen \cite{Cohen:1996ng} that massless two-flavor QCD 
at $T>T_c$ should be effectively symmetric under $\U(1)_A$ 
in the sense that two-point correlators in the $\pi$, $\sigma$, $\delta$,
and $\eta$ channels become all degenerate. This work was soon followed by 
counterarguments \cite{Evans:1996wf,Lee:1996zy}.  Later it was recognized 
by Laine and Veps\"{a}l\"{a}inen \cite{Laine:2003bd} that the 
$\U(1)_A$ violation in the flavor-singlet (axial) vector channel can be shown rigorously 
at least for high enough temperatures without any assumptions. 
This conclusion qualitatively agrees with \cite{Dunne:2010gd} that shows 
$\U(1)_A$ violation at high temperatures by calculating a nonzero 
splitting of scalar and pseudoscalar screening masses using a semiclassical 
dilute instanton gas picture \cite{Gross:1980br}.  
More recently, Aoki \emph{et al.}~\cite{Aoki:2012yj} claimed to have shown 
the effective restoration of the $\U(1)_A$ symmetry rigorously  
under certain assumptions, although the validity of their assumptions appears to be 
nontrivial and subtle (see Sec.~\ref{sec:comment}). 

Alongside these theoretical studies, the $\U(1)_A$ problem at finite temperature 
has also been studied intensively in first-principles lattice QCD 
simulations \cite{Bernard:1996iz,Chandrasekharan:1998yx}, but a consensus is 
not reached yet: effective \emph{restoration} of the $\U(1)_A$ 
symmetry was reported in simulations with overlap fermions \cite{Cossu:2013uua} and domain-wall 
fermions \cite{Bazavov:2012qja, Chiu:2013wwa, Cossu:2014aua,Tomiya:2014mma}, 
whereas a \emph{violation} of the $\U(1)_A$ symmetry was reported in 
simulations using staggered fermions \cite{Cheng:2010fe,Ohno:2011yr,Ohno:2012br,Dick:2015twa} 
and domain-wall fermions \cite{Buchoff:2013nra}.%
\footnote{In \cite{Dick:2015twa} the overlap Dirac operator was used to probe 
the Dirac spectra while configurations were generated with improved staggered fermions.}  
We warn that some of the simulations \cite{Bazavov:2012qja,Cheng:2010fe,Ohno:2011yr,Ohno:2012br,Buchoff:2013nra,Dick:2015twa} 
were performed for $2+1$ flavors; the effect of a heavier strange quark on the possible $\U(1)_A$ violation 
in the light-quark sector is not completely clear yet. 

In this paper, we do not try to solve this $\U(1)_A$ puzzle at finite temperature. 
Rather, we derive some rigorous results involving the $\U(1)_A$ symmetry in 
high-temperature QCD, \emph{assuming} that the $\U(1)_A$ symmetry is violated. 
(Since the $\U(1)_A$ violation must be present at least for high enough temperatures 
\cite{Laine:2003bd,Dunne:2010gd}, we consider its presence at \emph{all} $T>T_c$ to be quite plausible.) 
We first derive general expressions for chiral susceptibilities and the topological 
susceptibility at $T>T_c$ using the method of \cite{Chandrasekharan:1998yx,Kanazawa:2014cua}. 
They are used to highlight that, while the majority of the $\U(1)_A$ violation 
at \emph{small} volume comes from exact (topological) zero modes, 
the dominant contribution at \emph{large} volume comes from 
nonzero modes. We estimate finite-volume effects suffered by lattice QCD 
simulations with fixed topology, and propose a way to measure the $\U(1)_A$-violating effects 
reliably in a finite volume. Furthermore, we rigorously derive new sum rules and 
a new Banks--Casher-type relation for the Dirac eigenvalue spectra 
at $T>T_c$\,. This relation provides a link between 
the connected two-point correlation function of Dirac eigenvalues and the $\U(1)_A$ anomaly.
As a by-product of our spectral analysis, we find a remarkably simple proof of the 
Aoki--Fukaya--Taniguchi ``theorem'' on the effective restoration of the $\U(1)_A$ symmetry 
at high temperature, under the same assumptions as in \cite{Aoki:2012yj}. 
It should be emphasized that all our results are based on a systematic analysis of QCD. 
We expect that testing our exact relations and proposal in future lattice simulations 
should be a useful step towards the resolution of the $\U(1)_A$ puzzle at finite temperature.

The paper is organized as follows.
In Sec.~\ref{sec:review}, we review the argument of \cite{Laine:2003bd} 
for $\U(1)_A$ violation at high temperature. In Sec.~\ref{sec:topology}, 
we discuss the importance of topology and finite-volume effects on the breaking of the 
$\U(1)_A$ symmetry, as well as its implication for lattice QCD simulations. 
In Sec.~\ref{sec:Dirac}, we derive new spectral sum rules and a Banks--Casher-type relation 
for Dirac spectra concerning the $\U(1)_A$ anomaly. 
In Sec.~\ref{sec:comment}, we comment on the Aoki--Fukaya--Taniguchi ``theorem'' 
in \cite{Aoki:2012yj}. Section \ref{sec:conclusion} contains our conclusions.

In Appendix \ref{app:tensor}, we point out and correct the deficiency in the tensor decomposition 
of the anomalous correlation function in QCD at finite temperature studied by Itoyama and 
Mueller \cite{Itoyama:1982up}. In Appendix \ref{app:scaling}, we discuss the microscopic 
scaling that is different from \eqref{eq:rho_mic} in the main text.  
In Appendix \ref{ap:pois} we derive \eqref{eq:pois} in the main text.  
Throughout this paper, we will work on QCD with quarks 
in the fundamental representation of the gauge group $\SU(3)$.

\section{\boldmath $\U(1)_A$ anomaly at high temperature}
\label{sec:review}

In this section we review an argument for the $\U(1)_A$ anomaly 
in massless QCD at high temperature, given by 
Laine and Veps\"{a}l\"{a}inen \cite{Laine:2003bd}.%
\footnote{We thank M. Laine for explaining the logic of \cite{Laine:2003bd} in detail.}
Their argument is based on the anomaly relation 
\beq
  \label{anomaly}
  \der_{\mu} j^{A \mu} = \frac{N_f g^2}{32 \pi^2} 
  \epsilon^{\alpha \beta \mu \nu} G_{\alpha \beta}^a G_{\mu \nu}^a 
\eeq
for the axial current $j^{A \mu}=\bar \psi \gamma^{\mu} \gamma^5 \psi$, 
and the Debye screening of the gauge fields at high temperature. 
Here $N_f$ is the number of flavors, $g$ is the QCD coupling constant and 
$G_{\mu \nu}^a$ is the gluon field strength with $a$ being the color index. 

Let us work in Euclidean spacetime with the imaginary time $\tau = i t \in [0, \beta]$ and 
a spatial box of size $L_1\times L_2\times L_3$. We assume that both quarks and 
gauge fields obey periodic boundary conditions in spatial directions, whilst quarks 
(gauge fields) obey the anti-periodic (periodic) boundary condition in the temporal direction.   
As our interest is in the screening of gauge fields, we shall consider a spatial correlation function. 
Without loss of generality we choose a spatial separation in the $x_3$ direction.  
Then, for the axial ``charge'' (integrated over the volume transverse to the $x_3$ axis)
\beq
  Q^A_{3}(x_3) \equiv \int \dd \tau\,\dd^2 {\bm x}_\perp \, j^{A}_3 \,,
\eeq
one can show that $\akakko{Q^A_3(x_3)Q^A_3(y_3)}= 0$ 
in the thermodynamic limit if the axial symmetry is effectively restored, 
and $\akakko{Q^A_3(x_3)Q^A_3(y_3)}\ne 0$ if it is broken.   
Here $\akakko{\dots}$ is a statistical average, and the coordinates $x_3$ and $y_3$ 
are entirely arbitrary.  The fact that $\akakko{Q_3(x_3) Q_3(y_3)}=0$ when $\U(1)_A$ is 
unbroken may be shown in two steps as follows. 
Suppose we add a constant external field $a$ that couples to the axial charge $Q_3^A$. 
Then the action acquires an additional contribution 
$\displaystyle \delta S = a \int \dd x_3\ Q_3^A(x_3)$. 
This leads to the susceptibility
\ba
	\chi^A = \frac{1}{V_4}\frac{\der^2}{\der a^2}\log Z(a)\Big|_{a=0} & = 
	\frac{1}{V_4}\akakko{\int \dd x_3\ Q_3^A(x_3)\int \dd y_3\ Q_3^A(y_3)}
	\\
	& = \frac{1}{\beta L_1 L_2} \int \dd x_3 \akakko{Q_3^A(x_3)Q_3^A(0)} , 
\ea
where $V_4=\beta L_1L_2L_3$ is the total volume of spacetime. 
As $\U(1)_A$ is conserved by assumption, $\chi^A$ must be finite in the 
thermodynamic limit; hence $\chi^A<\infty$ as $L_3\to \infty$. 
This completes the first step of the proof. 

In the second step, we use the fact that $Q_3^A(x_3)$ is independent of 
$x_3$. Actually, this is a direct consequence of the (assumed) conservation of 
the axial current:
\ba
	\frac{\dd}{\dd x_3}Q_3^A(x_3) & =  
	\int \dd \tau\,\dd^2 {\bm x}_\perp \, 
	\der_3 j^{A}_3
	\\
	& = - \int \dd \tau\,\dd^2 {\bm x}_\perp 
	\mkakko{ \der_0 j^A_0 + \nabla_\perp \cdot j_\perp^A } =0\,,
	\label{eq:Q3in}
\ea
where the last step follows from the boundary conditions for fields. 
Note that \eqref{eq:Q3in} holds rigorously in a finite volume. 
Then it is clear that we must have
\ba
	\frac{\dd}{\dd x_3}\akakko{Q_3^A(x_3)Q_3^A(0)} & = 
	\akakko{\frac{\dd}{\dd x_3}Q_3^A(x_3)Q_3^A(0)}=0 \,.
	\label{eq:q3q3}
\ea
Combining \eqref{eq:q3q3} with the finiteness of $\chi^A$ in the limit 
$L_3\to \infty$, we conclude that 
$\akakko{Q_3^A(x_3)Q_3^A(0)}$ must vanish in the thermodynamic limit for 
an arbitrary $x_3$. This completes the second step of the proof. 

Next, recall that the anomaly relation (\ref{anomaly}) can be rewritten as a total derivative, 
\ba
  \der_{\mu} j^{A}_{\mu} = \der_{\mu} K_{\mu}, \qquad
  K_{\mu} = \frac{N_f g^2}{8 \pi^2} \epsilon_{\mu \nu \lambda \rho}
	\mkakko{
  		A_{\nu}^a \der_{\lambda} A_{\rho}^a + 
  		\frac{g}{3}f^{abc}   A_{\nu}^a A_{\lambda}^b A_{\rho}^c
	}.
\ea
In terms of this $K_\mu$, the so-called Chern-Simons charge reads as
\beq
  Q^{\rm CS}_3(x_3) \equiv \int \dd\tau \, 
  \dd^2 {\bm x}_{\perp} \, K_3\,. 
\eeq
Because the gauge fields are Debye screened at high temperature, 
the correlation of $Q^{\rm CS}_3$ should decay exponentially:
\beq
  \label{Debye}
  \langle Q^{\rm CS}_3(x_3) Q^{\rm CS}_3(y_3)  \rangle 
  \sim \ee^{-|x_3 -y_3|/\xi} 
\eeq
at $|x_3 - y_3| \gg \xi$, where $\xi$ is the correlation length (or the inverse 
Debye mass); $\xi^{-1} \approx gT$ at leading order in $g$. 
Although the correlator \eqref{Debye} may appear to vanish trivially due to 
the non-gauge-invariance of $Q^{\rm CS}_3$, this is not necessarily true. 
As $Q^{\rm CS}_3$ is gauge invariant up to surface terms, it becomes 
gauge invariant and its correlator becomes nonzero once we fix boundary 
conditions for the gauge fields. 

Since $Q_3^A$ and $Q^{\rm CS}_3$ have the same quantum numbers, 
$Q^{\rm CS}_3$ contributes to the correlator of $Q_3^A$.  
Hence we expect
\ba
  \langle Q^{A}_3(x_3) Q^{A}_3(y_3) \rangle 
  & \propto \langle Q^{\rm CS}_3(x_3) Q^{\rm CS}_3(y_3)  \rangle 
  \sim \ee^{-|x_3-y_3|/\xi} \,.
  \label{QA3QA3}
\ea
This shows that $\langle Q^{A}_3(x_3) Q^{A}_3(y_3)\rangle$ 
cannot be a constant as a function 
of $x_3$, indicating that the $\U(1)_A$ symmetry is certainly violated 
at the level of correlation functions of quark bilinears in QCD 
at sufficiently high temperature. 

It would then be quite natural to expect, by continuity, 
that the $\U(1)_A$ symmetry be violated 
at \emph{any} $T>T_c$\,.  
In the following, we shall assume $\U(1)_A$ violation for 
$T>T_c$ in the thermodynamic limit and pursue its consequences in detail.

\section{Topology and finite-volume effect}
\label{sec:topology}

In this section we write down the most general QCD partition function for $T>T_c$
in terms of quark masses and derive general expressions for the chiral susceptibilities 
and topological susceptibility based on the method of 
\cite{Chandrasekharan:1998yx,Kanazawa:2014cua}. 
Our arguments here are based on symmetries of 
QCD and a systematic expansion in terms of a small parameter $m/T \ll 1$, 
and are fully under theoretical control. We then elucidate the contributions of 
topology and finite-volume effects to the violation of the $\U(1)_A$ symmetry
(characterized by the difference of two-point functions $\chi_{\pi} - \chi_{\delta}$
to be defined below), and discuss possible implications for lattice QCD simulations.
For definiteness, we will concentrate on two-flavor QCD below. 

\subsection{Partition function and topological susceptibility}

We consider the partition function of two-flavor QCD at finite temperature as a function of 
quark masses $m_{u,d}$. Since there are no massless modes at $T>T_c$ in the chiral limit,%
\footnote{In the imaginary-time formalism, contributions of massless quarks is infrared (IR) 
finite because the lowest Matsubara frequency for fermions $\sim \pi T$ acts 
as an effective IR cutoff. 
}
the free energy density should be \emph{analytic} in quark mass.%
\footnote{It should be stressed that analyticity of the partition function breaks down if the 
zero-temperature part is thrown away. For example, in a non-interacting theory 
the free energy of quarks after subtraction of the zero-temperature part includes  
a term $\sim m^4\log(m/\pi T)$ \cite{Kapusta:2006pm}, which is not 
analytic in $m$. 
}

To write down the general form of the free energy, 
we consider a generic quark mass matrix $M$ ($= 2\times 2$ matrix in the flavor space) 
and let $M$ transform under the symmetry, 
${\cal G} \equiv \SU(2)_{R} \times \SU(2)_{L} \times \U(1)_A$, 
so that the quark mass term in the QCD Lagrangian 
\beq
  {\cal L}_{\rm mass} = \psi_{L}^{\dag} M \psi_R + {\rm h.c.} 
\eeq
is invariant under ${\cal G}$. Here $\psi_{R,L}$ are the right- and left-handed quarks,
which transform under ${\cal G}$ as 
$\psi_R \rightarrow \ee^{i\theta_A} V_R \psi_R$ and 
$\psi_L \rightarrow \ee^{-i\theta_A} V_L \psi_L$, 
respectively, where $\theta_A$ is a $\U(1)_A$ rotation angle  
and $V_{R, L}\in\SU_{R, L}(2)$. It then follows that $M$ 
should transform as 
$M \rightarrow \ee^{-2 i\theta_A} V_L M V_R^{\dag}$. 

Noting that the free energy density at $T>T_c$ is invariant 
under the restored $\SU(2)_R \times \SU(2)_L$ chiral symmetry but 
\emph{not} under the $\U(1)_A$ symmetry,  
the partition function of QCD in a spatial volume $V_3$ can be expanded 
in terms of a small parameter $m_{u,d}/T \ll 1$ as 
\cite{Chandrasekharan:1998yx,Kanazawa:2014cua} 
\ba
	  Z(T,V_3,M) & = \exp\left[ - \frac{V_3}{T} f(T,V_3,M) \right]\,,
	  \label{eq:Z_ini_}
	  \\
	  f(T,V_3,M) & = f_0 - f_2 \tr M^\dagger M 
	  - f_A (\det M+\det M^\dagger) + \calO(M^4)\,,  
	  \label{eq:Zex}
\ea
where $f_0, f_2$ and $f_A$ are functions of $T$ and $V_3$. 
We assume that this expansion has a nonzero radius of convergence. 
The term $\propto f_A$ represents the effect of axial anomaly: 
for a $\U(1)_A$ rotation $\psi \rightarrow \ee^{i \gamma_5 \theta_A} \psi$, this term transforms as 
$\det M \rightarrow \ee^{4i \theta_A} \det M$, so it breaks $\U(1)_A$ down to $\mathbb{Z}_{4}$. 
The absence of $\calO(M)$ terms is consistent with the vanishing chiral condensate in the chiral limit for $T>T_c$\,. 
In the following we will disregard the $\calO(M^4)$ terms in the free energy as they are suppressed 
by additional powers of $m_{u,d}/T\ll1$. Since the partition function (\ref{eq:Z_ini_}) is obtained with a systematic
expansion, this will be called the ``effective theory'' in this paper (although
there is no dynamical field in it).

We now turn to the study of topological sectors. 
As is well known, the $\theta$ angle can be incorporated into  
the partition function via $M\to M \ee^{i\theta/N_f}$ 
\cite{Leutwyler:1992yt}, where $N_f=2$ is of our interest here.   
Then the partition function in a sector of given topological charge 
\ba
	\Q \equiv  \frac{g^2}{32\pi^2} \int \dd^4x\,  G^a_{\mu \nu} \tilde G^a_{\mu \nu}
	\label{eq:defQ}
\ea
is obtained, from \eqref{eq:Zex}, as 
\ba
  Z_{\Q} (T,V_3,M) & \equiv \oint \frac{\dd\theta}{2\pi} 
  \ee^{-i\Q \theta} Z(T,V_3,M \!\ee^{i\theta/2}). 
  \\
  & = \ee^{-V_4 [f_0-f_2(m_u^2+m_d^2)]}
  \oint \frac{\dd\theta}{2\pi}  \ee^{-i\Q \theta}
  \ee^{ 2V_4 f_Am_um_d\cos\theta}
  \\
  & = \ee^{-V_4 [f_0-f_2(m_u^2+m_d^2)]}  I_{\Q}(2V_4 f_Am_um_d)\,,
  \label{eq:Znu}
\ea
where $V_4\equiv V_3/T$ is the spacetime volume, 
$I_\Q$ is the modified Bessel function of $\Q$-th order, 
and $M=\diag(m_u,\,m_d)$ was substituted. 
Intriguingly, the probability distribution of $\Q$ is 
proportional to $I_\Q$ in one-flavor QCD, too \cite{Leutwyler:1992yt}.%
\footnote{An analogous toy model was also studied in \cite[Appendix A]{Aoki:2007ka}.
}

The Taylor expansion of \eqref{eq:Znu} in powers of quark masses starts 
with $(V_4 f_Am_um_d)^{|\Q|}$, which is the contribution of exact zero modes. 
Hence the topological sectors with $\Q\ne 0$ will all drop out in the chiral limit if 
$V_4$ is finite. By contrast, topological fluctuations will not be suppressed at all 
even near the chiral limit if $V_4$ is sufficiently large. This subtle balance 
between topology and volume has an important practical consequence for lattice simulations, 
as we will discuss shortly. 

An important quantity that characterizes topological fluctuations is the 
mean square of the topological charge at $\theta=0$,
\ba
  \langle \Q^2 \rangle = 
  \sum_{\Q=-\infty}^{\infty} \Q^2 \frac{Z_\Q}{Z}
  = 2V_4 f_Am_um_d\,,
  \label{eq:nusquared}
\ea
where (\ref{eq:Znu}) was used. 
The topological susceptibility is then given 
\cite{Chandrasekharan:1998yx,Kanazawa:2014cua} by 
\beq
  \label{chi}
  \chi_{\rm top} \equiv \frac{\langle \Q^2 \rangle}{V_4} 
  = 2f_Am_um_d.
\eeq 
Alternatively, one can reach \eqref{chi} by considering that, 
with the replacement $M\to M \ee^{i\theta/2}$ in (\ref{eq:Zex}), the 
$\theta$-dependence of the free energy reads
\beq
  \label{eq:f}
  f(\theta) =  \tilde f- 2 f_A m_u m_d \cos \theta + \calO(m^4),
\eeq
where $\tilde f \equiv f_0 - f_2 (m_u^2 + m_d^2)$ is the term independent of $\theta$. 
The topological susceptibility is then
\beq
  \chi_{\rm top} \equiv \left. 
  \frac{\der^2 f(\theta)}{\der \theta^2} \right|_{\theta=0} 
  = 2 f_Am_um_d\,,
\eeq
which is the same as (\ref{chi}).
Note that our result, obtained assuming $f_A\ne 0$, does not agree with the 
result by Aoki \emph{et al.} \cite{Aoki:2012yj} that $\langle \Q^2\rangle/V_3\to 0$ as
 $V_3\to \infty$ for a small but nonzero $m$.

The topological susceptibility here should be contrasted with that of the QCD vacuum,
\beq
  \label{chi_SB}
  \frac{\langle \Q^2 \rangle}{V_4} =    
  \Sigma(m_u^{-1}+m_d^{-1})^{-1} 
\eeq
with $\Sigma$ being the magnitude of the chiral condensate \cite{Leutwyler:1992yt}.
This difference can be understood in the following way. In the 
presence of chiral symmetry breaking, the quark mass dependence of the 
free energy can be expanded in terms of the quark mass as
\beq
  \label{exp}
  f = f_0 - \Sigma [\tr (M U^{\dag}) + \tr (M^{\dag} U)] 
  + \calO(M^2),
\eeq
where $U$ denotes the $\SU(2)$ Nambu-Goldstone field associated with 
spontaneous chiral symmetry breaking. 
The topological susceptibility is dominated by the second term in (\ref{exp}) and is given by (\ref{chi_SB}), 
with the higher order contributions being $\calO(M^2)$.
When chiral symmetry is restored $(\Sigma = 0)$, on the other hand, the leading
contribution to the topological susceptibility is $\calO(M^2)$ and is given by (\ref{chi}). 
The behavior $\chi_{\rm top} \propto m_u m_d$ is also found 
in the 2SC phase of dense QCD for the same reason \cite{Son:2001jm}. 

From the partition function (\ref{eq:Zex}), higher moments of $\Q$ 
can also be derived \cite{Kanazawa:2014cua} as 
\begin{subequations}
\label{eq:higher}
\begin{align}
  \langle \Q^2 \rangle  & = {\cal A},
  \\
  \langle \Q^4 \rangle  & = {\cal A}(1+3{\cal A}),
  \\
  \langle \Q^6 \rangle  & = {\cal A}(1+15{\cal A}+15{\cal A}^2),
  \\
  \langle \Q^8 \rangle  & = {\cal A}(1+63{\cal A}+210{\cal A}^2+105{\cal A}^3),
\end{align}
\end{subequations}
with ${\cal A} \equiv 2V_4 f_A m_u m_d$\,, while all odd moments vanish.

\subsection
[Chiral susceptibilities and $\U(1)_A$ anomaly]
{\boldmath Chiral susceptibilities and $\U(1)_A$ anomaly}
\label{sc:CSanom}

We now turn to two-point correlation functions of quark bilinears 
(also called chiral susceptibilities). 
The utility of chiral susceptibilities in two-flavor QCD as a convenient 
probe for the $\U(1)_A$ anomaly has been advocated long time ago  
\cite{Shuryak:1993ee,Cohen:1996ng}, and nowadays they are measured in 
lattice simulations with dynamical quarks 
\cite{Bazavov:2012qja,Cossu:2013uua,Buchoff:2013nra}. 
It is therefore of primary interest to relate these susceptibilities 
to the coefficients $f_0$, $f_2$ and $f_A$ of the effective theory 
\eqref{eq:Zex} \cite{Chandrasekharan:1998yx}. 
In this paper we will work with the following definitions for chiral susceptibilities:%
\begin{subequations}
	\begin{align} 
	  \chi_\sigma & 
	  \equiv \int \dd^4x \Big[
	    \langle \bar \psi \psi(x)\bar\psi\psi(0) \rangle 
	    - \langle \bar \psi\psi \rangle^2 
	  \Big] \,,
	  \\
	  \chi_\pi & 
	    \equiv \int \dd^4x \Big[
	    \langle \bar \psi i\gamma_5 \tau_3\psi(x)
	    \bar\psi i\gamma_5 \tau_3\psi(0)
	    \rangle - \langle \bar \psi i\gamma_5 \tau_3
	    \psi \rangle^2 
	  \Big]\,, 
	  \label{eq:chi_d}
	  \\
	  \chi_\delta & 
	  \equiv \int \dd^4x \Big[
	    \langle \bar \psi\tau_3\psi(x)\bar\psi\tau_3\psi(0) \rangle 
	    - \langle \bar \psi\tau_3\psi \rangle^2
	  \Big]  \,,
	  \label{eq:chidelta}
	  \\
	  \chi_\eta & 
	  \equiv \int \dd^4x \Big[
	    \langle \bar \psi i\gamma_5 \psi(x)\bar\psi i \gamma_5 \psi(0) \rangle 
	    - \langle \bar \psi i\gamma_5 \psi \rangle^2 
	  \Big]  \,,  
	\end{align}
	\label{eq:chis}%
\end{subequations}
where $\bar\psi\psi=\bar u u +\bar d d$ and 
$\bar \psi i \gamma_5 \psi = \bar u i \gamma_5 u + \bar d i \gamma_5 d$ 
in our notation, and $\tau_3$ is the third Pauli matrix. 
Note that these definitions do not necessarily coincide with those in the literature. 

Inserting $M=\diag(m_u,m_d)$ into $f(T,V_3,M)$, we obtain 
\ba
  \chi_\sigma 
  & = \frac{1}{V_4}\left(
    \frac{\der^2}{\der m_u^2} + 
    \frac{\der^2}{\der m_d^2} + 
    2 \frac{\der^2}{\der m_u \der m_d}
    \right)\log Z =  4f_2  + 4f_A 
  \label{eq:ch1}
  \\
  \chi_\delta & = \frac{1}{V_4}\left(
    \frac{\der^2}{\der m_u^2} + 
    \frac{\der^2}{\der m_d^2} - 
    2 \frac{\der^2}{\der m_u \der m_d}
  \right)\log Z
  =  4f_2  - 4f_A 
  \label{eq:ch3}
\ea
up to $\calO(m^2)$ corrections.  
Similarly, inserting $M=\diag(m_u+ib,m_d-ib)$ into $f(T,V_3,M)$ we find
\ba
    \chi_\pi & = \frac{1}{V_4}
  \left[ \frac{1}{Z}\frac{\der^2 Z}{\der b^2}\Big|_{b= 0} 
  - \left({\frac{1}{Z}\frac{\der Z}{\der b}\Big|_{b= 0}}\right)^{\! \! 2} 
  \right]
  = 4f_2 + 4f_A \,. 
  \label{eq:ch2}
\ea
In the same way one can also show $\chi_\eta=4f_2-4f_A$. 
The equality $\chi_\sigma=\chi_\pi$, as well as 
$\chi_\delta=\chi_\eta$, is a direct consequence 
of the restored $\SU(2)_R\times \SU(2)_L$ chiral symmetry.   

The axial anomaly manifests itself in the difference \cite{Chandrasekharan:1998yx} 
\ba
    \chi_\pi-\chi_\delta &=  8 f_A  +\calO(M^2) \,,
    \label{eq:c-dif}
\ea
where the correction is displayed for completeness. 
Practically, the correlators of $\pi$ and $\delta$ are most convenient in lattice simulations, 
because they have no disconnected components (for degenerate masses). 
Note that, if one wishes, it is entirely straightforward to extend the calculation for 
\eqref{eq:c-dif} to higher orders by including $\U(1)_A$-violating 
terms such as $(\det M)^2$ and $(\tr MM^\dagger)(\det M)$ in the free energy \eqref{eq:Zex}. 
However this is not expected to bring about quantitative differences when $M/T\ll 1$. 

In \cite{Dunne:2010gd}, the two-point correlators $\langle\pi(x)\pi(0)\rangle$ and 
$\langle\delta(x)\delta(0)\rangle$   
(rather than their spatial integrals, $\chi_\pi$ and $\chi_\delta$) were calculated directly in high-temperature 
QCD by using the 't Hooft vertex of instantons. They found that the difference of the two correlators 
decreases at high temperature but does \emph{not} vanish exactly, in agreement with the general 
argument presented in Sec.~\ref{sec:review}.  

It was emphasized in \cite{Evans:1996wf,Lee:1996zy} that the dominant contribution to 
$\chi_\pi-\chi_\delta$ comes from exact zero modes in the $\Q=\pm 1$ sector. 
A more recent paper \cite{Aoki:2012yj} argues to the contrary that contributions of 
exact zero modes is suppressed in the thermodynamic limit. In what follows we aim 
to clarify this issue. 

Let us first decompose the anomalous contribution \eqref{eq:c-dif} into 
contributions from each topological sector. We assume $\theta=0$ in the following. 
Since the second terms in \eqref{eq:chi_d} and \eqref{eq:chidelta}   
vanish for degenerate masses, it follows that
\ba
  \lim_{m_{u,d}\to m}(\chi_\pi-\chi_\delta) & = 
  \int \dd^4x \Big[
    \langle \bar \psi i\gamma_5 \tau_3\psi(x)\bar\psi i\gamma_5  
    \tau_3\psi(0) \rangle - \langle \bar  
    \psi\tau_3\psi(x)\bar\psi\tau_3\psi(0) \rangle 
  \Big]
  \nonumber 
  \\ & 
  = 
  \frac{1}{V_4}\left( \frac{1}{Z}
  \frac{\der^2 Z}{\der b^2}\Big|_{b= 0}
  - \frac{1}{Z}\frac{\der^2 Z}{\der c^2}\Big|_{c= 0} \right)
  \label{eq:chi_diff}
  \\
  & \equiv \sum_{\Q=-\infty}^{\infty}\frac{Z_\Q}{Z}P_\Q\,,
  \label{eq:PP}
\ea
where it is tacitly assumed in \eqref{eq:chi_diff} that 
the first term is evaluated for $M=\diag(m+ib,m-ib)$ and the second term for $M=\diag(m+c,m-c)$.  
In \eqref{eq:PP} we defined the contribution $P_\Q$ from the sector of topological 
charge $\Q$ as
\ba
  P_\Q & \equiv \frac{1}{V_4}\left(\frac{1}{Z_\Q}
  \frac{\der^2 Z_\Q}{\der b^2}\Big|_{b=0}
  - \frac{1}{Z_\Q}\frac{\der^2 Z_\Q}{\der c^2}\Big|_{c=0} 
  \right)
  \label{Pnu}
  \\
  & = \left[ 4f_2 + 4f_A \frac{I'_\Q(2V_4 f_A m^2)}{I_\Q(2V_4 f_A m^2)} \right] 
  - \left[ 4f_2 - 4f_A \frac{I'_\Q(2V_4 f_A m^2)}{I_\Q(2V_4 f_A m^2)} \right]
  \\
  & = 8f_A \frac{I'_\Q(2V_4 f_A m^2)}{I_\Q(2V_4 f_A m^2)}\,. 
\ea
Using the identity $I'_\Q(x) = \frac{\Q}{x} I_\Q(x) + I_{\Q + 1}(x)$ or 
$I'_\Q(x) = - \frac{\Q}{x} I_\Q(x) + I_{\Q - 1}(x)$ depending on 
the sign of $\Q$, one may cast $P_\Q$ into a suggestive form
\ba
  P_\Q = \left\{
  \begin{array}{ll}
    \displaystyle 
    \frac{4}{V_4m^2}\Q + 8f_A \frac{I_{\Q+1}(2V_4 f_A m^2)}{I_\Q(2V_4 f_A m^2)} 
    & \quad \text{for}~\Q\geq 0\,,
    \\
    \displaystyle 
    \frac{4}{V_4m^2} |\Q| + 8f_A \frac{I_{\Q-1}(2V_4 f_A m^2)}{I_\Q(2V_4 f_A m^2)}
    & \quad \text{for}~\Q<0 \,. 
  \end{array} 
  \right. 
  \label{Pnu2}
\ea
The first terms in \eqref{Pnu2} are the contributions from exact zero modes. 
This can be easily seen by plugging $Z_\Q \propto (m^2+b^2)^{|\Q|}$ and 
$Z_\Q\propto (m^2-c^2)^{|\Q|}$ 
into the first and the second terms in \eqref{Pnu}, respectively. 
Therefore the $\U(1)_A$-violating contribution \eqref{eq:PP} may be split 
into the zero-mode fraction%
\footnote{It is intriguing that \eqref{eq:Rz} below has exactly the same form as 
the fraction of zero modes for the chiral condensate in one-flavor QCD 
\cite[eq.~(7.3)]{Leutwyler:1992yt}, under the identification 
$2V_4f_Am^2 \leftrightarrow V_4\Sigma m$\,.
} 
and the nonzero-mode fraction as
\ba
	  \lim_{m_{u,d}\to m}(\chi_\pi-\chi_\delta) =  8 f_A (S_{z} + S_{nz}) \,, 
\ea
where
\ba
	  S_z & \equiv \frac{1}{8f_A} \bigg( 
	  	2 \sum_{\Q=1}^{\infty}\frac{Z_\Q}{Z}\frac{4}{V_4m^2}\Q 
	  \bigg) 
	  = \ee^{-2V_4 f_A m^2}\big[ I_0(2V_4 f_A m^2)+I_1(2V_4 f_A m^2)\big]\,,
	  \label{eq:Rz}
	  \\
	  S_{nz} & = 1 - S_z\,. 
\ea  
\begin{figure}[t!]
  \begin{center}
    \includegraphics[width=9cm,clip]{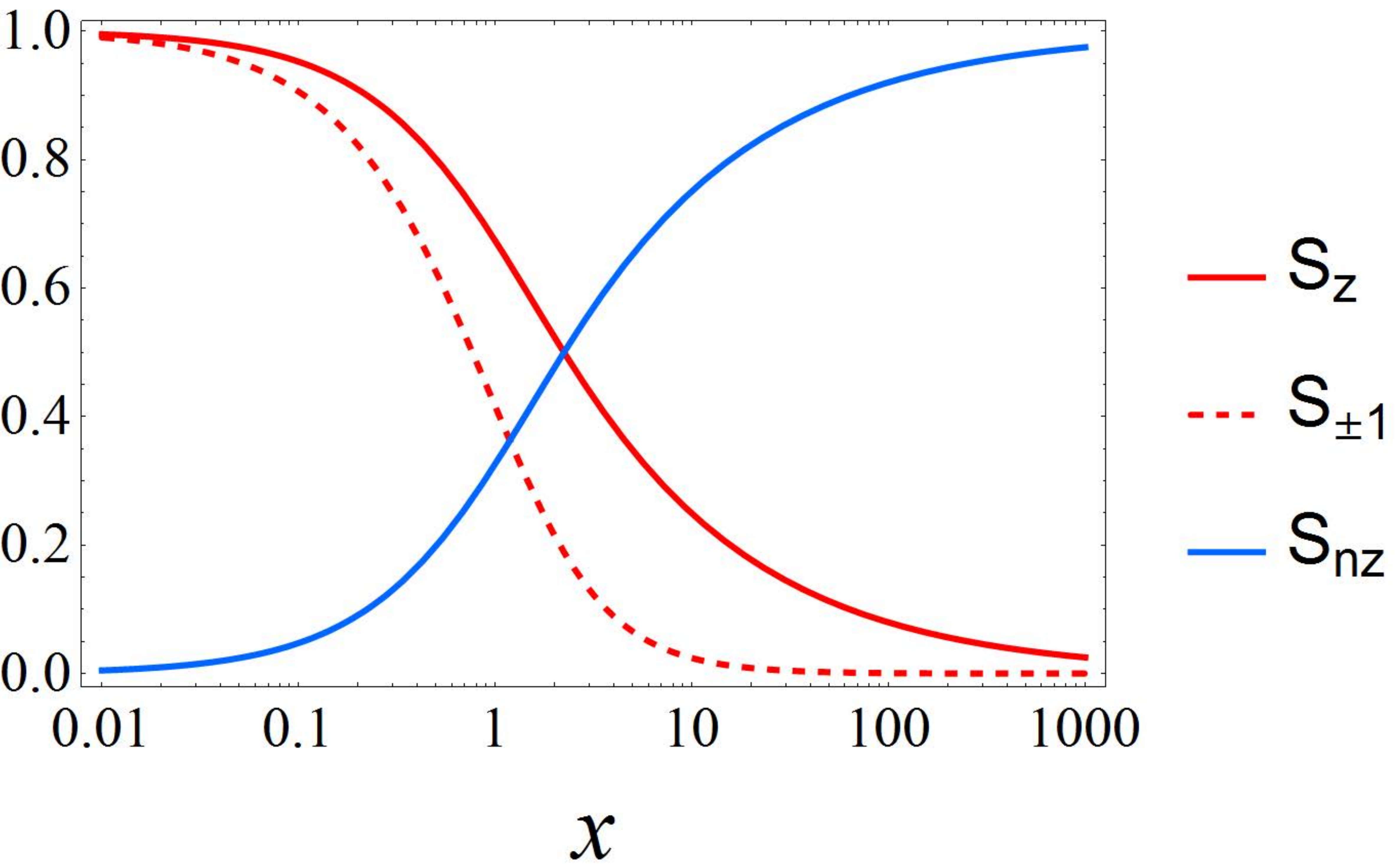}
  \end{center}
  \vspace{-1.5\baselineskip}
  \caption{\label{fig:Rs} 
		Relative contributions of zero and nonzero Dirac eigenmodes
		to $\chi_\pi-\chi_\delta$ as a function of $x\equiv 2V_4 f_A m^2$.
  }
\end{figure}%
In addition, the contribution of the $\Q=\pm 1$ sectors to $S_z$ is defined as 
\ba
  S_{\pm 1} & \equiv 
  \frac{1}{8f_A} \left(2 \frac{Z_1}{Z}\frac{4}{V_4 m^2}\right)
  = \frac{1}{V_4f_Am^2}\ee^{-2V_4f_Am^2}I_1(2V_4f_Am^2)\,. 
\ea
The quantities $S_z$, $S_{nz}$ and $S_{\pm 1}$ are plotted in Figure \ref{fig:Rs} 
as functions of $x\equiv 2V_4 f_A m^2$. 

We observe that, in a small volume or near the chiral limit ($x\ll 1$), $\chi_\pi-\chi_\delta$ 
is dominated by the contribution of exact zero modes in the $\Q=\pm 1$ sector, 
as argued in \cite{Evans:1996wf,Lee:1996zy}. 
By contrast, if we take the thermodynamic limit ($x\gg 1$), the contribution of 
nonzero modes dominates, and the exact zero modes are completely \emph{irrelevant}. 
This can be understood from \eqref{eq:nusquared}: since 
$\langle\Q^2\rangle \sim V_4 f_A m^2$, 
one naturally expects $\langle|\Q|\rangle =\calO(\sqrt{V_4})$, implying that the first term in \eqref{Pnu2} 
is suppressed in a large volume.%
\footnote{In this inspection, the positivity of the path-integral measure 
plays an essential role. We note that the suppression of exact zero modes 
does not hold in general for negative or complex masses 
\cite{Leutwyler:1992yt,Kanazawa:2011tt,Verbaarschot:2014upa}.}
On the other hand, the second  term in \eqref{Pnu2} 
tends to $8f_A$, which is the same value as in the full theory \eqref{eq:c-dif}. 
This means that the anomaly $(f_A\ne 0)$ in the thermodynamic limit must be attributed to  
\emph{nonzero} Dirac eigenmodes.    
The $\Q=\pm 1$ sector does not play a distinguished role. Indeed,  
one can show for $x\gg 1$ that  
$Z_{\Q}/Z$ obeys a Gaussian distribution (see also \cite{Leutwyler:1992yt}), 
according to which $Z_{\Q}/Z\sim 1/\sqrt{V_4 f_A m^2}$ for $|\Q|\lesssim 
\sqrt{V_4 f_A m^2}$ and is suppressed otherwise. 
Therefore, if the volume is sufficiently large with a fixed nonzero mass, 
all contributions to $\chi_\pi-\chi_\delta$ from the sectors with 
$|\Q|\lesssim \sqrt{V_4 f_A m^2}$ are \emph{equally important}, 
in contradistinction to the finite-volume regime $(x\lesssim 1)$
where only the $\Q=\pm 1$ sectors contribute to $\chi_\pi-\chi_\delta$\,. 

To avoid confusion, we stress that the total amount of $\chi_\pi-\chi_\delta$ is equal to
$8f_A$ irrespective of the value of $x$; the order-of-limit issue does not arise, 
of course, because there is no long-range-order in QCD above $T_c$\,. The reason 
the exchange of dominance occurs between zero and nonzero modes as we vary 
the volume is that a long-range correlation is induced once the 
global topological charge is fixed \cite{Leutwyler:1992yt}.

\subsection{Implications for lattice QCD simulations}

We now discuss implications of the above results for lattice QCD simulations. 
So far the $\U(1)_A$ anomaly at high temperature has been thoroughly investigated 
on the lattice (as reviewed in Sec.~\ref{sc:intro}), but despite efforts, 
a definitive conclusion on the (non-)restoration of the $\U(1)_A$ symmetry is not reached yet. 
This is not surprising, considering that the physics of $\U(1)_A$ anomaly 
is highly sensitive to the explicit breaking of chiral symmetry by lattice discretization; 
even domain-wall fermions have serious problems, as pointed out in \cite{Tomiya:2014mma}. 
In this regard, the most reliable simulations are those in \cite{Cossu:2013uua} employing dynamical 
overlap fermions. They reported restoration of the $\U(1)_A$ symmetry based on simulations 
with a fixed global topological charge ($\Q=0$). They also evaluated possible finite-size effects 
associated with the topology fixing, by using the formalism developed in \cite{Brower:2003yx,Aoki:2007ka}.  
Here we wish to revisit this issue based on our effective-theory framework.  
\begin{figure}[t!]
  \begin{center}
    \includegraphics[width=7cm,clip]{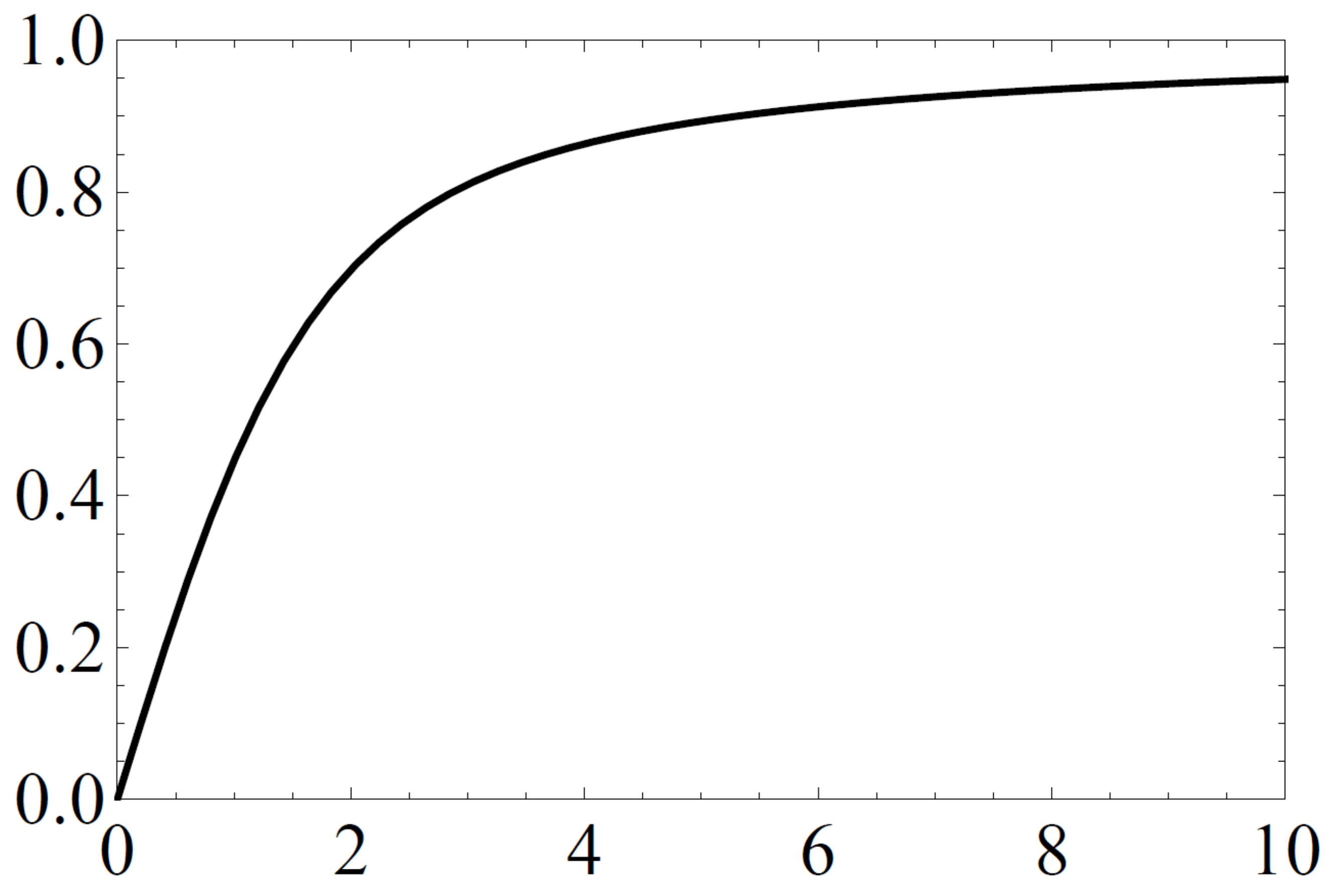}
    \put(-98,-12){\Large $x$}
    \put(-240,68){\Large $\displaystyle \frac{I_1(x)}{I_0(x)}$}
    \vspace{-\baselineskip}
  \end{center}
  \caption{The magnitude of $(\chi_\pi-\chi_\delta)\big|_{\Q=0}$ 
  normalized by $(\chi_\pi-\chi_\delta)\big|_{\rm full}$ as a function of 
  $x=2V_4f_Am^2$. At large volume $(x\gg 1)$, $I_1(x)/I_0(x)\simeq 1-\frac{1}{2x}$.}
  \label{fig:fract}
\end{figure}

It follows from \eqref{eq:PP} that in the topologically trivial sector ($\Q=0$) we have  
\ba
  \label{eq:chichi0}
  \chi_\pi-\chi_\delta = 8f_A \frac{I_1(2V_4f_Am^2)}{I_0(2V_4f_Am^2)}\,.
\ea
The ratio of \eqref{eq:chichi0} to $(\chi_\pi-\chi_\delta)\big|_{\text{full}} = 8f_A$ 
is plotted in Figure \ref{fig:fract}. It shows that the ratio tends to $0$ for small $x$ 
and obscures the nonzero value in the full theory. This signals a strong finite-volume 
effect at small $x$. It seems necessary to ensure at least 
$x=2V_4f_Am^2\gtrsim 1$ in order to 
observe a nonzero value of $\chi_\pi-\chi_\delta$ clearly. 

Our result so far is rigorous, as long as $f _A \neq 0$ and 
the $\calO(M^4)$ correction to \eqref{eq:Zex} can be neglected. 
At sufficiently high temperature $T\gg T_c$ we may resort to the dilute instanton gas 
approximation \cite{Gross:1980br}, which yields 
\beq
  f_A \sim T^2 \ee^{-8\pi^2/g^2} \sim 
  T^2(\Lambda/T)^{(11N_c-2N_f)/3} \propto T^{-23/3} 
  \label{eq:fDIGA}
\eeq
for $N_c = 3$ and $N_f = 2$. Since it decays so rapidly, 
it would be a challenging task to achieve a sufficiently large volume 
that satisfies $2V_4f_Am^2 \gtrsim 1$ while keeping $m$ small. 
On the other hand, near $T_c$\,, the asymptotic formula \eqref{eq:fDIGA} breaks down and 
we do not exactly know how small $f_A$ is.   

One way to extract $f_A$ from a topology-fixed simulation is as follows: 
if the simulation volume $V_4$ is not large and hence $V_4f_Am^2\lesssim 1$, 
then we have, to a good approximation, 
\ba
  (\chi_\pi-\chi_\delta)\Big|_{\Q=0} \simeq 
  8V_4f_A^2m^2 \,
  \label{eq:chichi00}
\ea
from \eqref{eq:chichi0}, where we used $I_1(x)/I_0(x)\simeq x/2$ for $x\lesssim 1$. 
Therefore in principle one can extract $f_A$ 
by fitting the lattice data to the formula \eqref{eq:chichi00}. 
This is a proposal for future lattice simulations with 
overlap fermions.

\section{\boldmath 
Dirac eigenvalue spectra and $\U(1)_A$ anomaly}
\label{sec:Dirac}

In this section we derive new spectral sum rules and a novel Banks--Casher-type relation 
which link the Dirac spectrum to the violation of the $\U(1)_A$ symmetry 
in high-temperature QCD. We then discuss possible forms of the spectral functions. 
Throughout this section we will focus on two-flavor QCD. 
We denote the purely imaginary eigenvalues of the Euclidean 
Dirac operator $D = \gamma_{\mu}(\partial_{\mu}+ig A_{\mu})$ 
by $\{ i\lambda_n \}_n$ with $\lambda_n\in\RR$ 
and define the spectral density for a fixed gauge field $A_\mu$ in a finite volume as
\ba
	\rho^A(\lambda) & \equiv \sum_{n}\delta(\lambda-\lambda_n)\,.  
	\label{eq:rhodef}
\ea
Chiral symmetry of the Dirac operator, $\{D,\gamma_5\}=0$, 
ensures $\rho^A(\lambda)=\rho^A(-\lambda)$.

\subsection{Spectral sum rules I:~macroscopic limit}
\label{sec:sum_rules}

The partition function of two-flavor QCD in the sector of topological charge 
$\Q$ is given by
\ba
  Z_\Q & = \bigg\langle 
  {\prod_k} ( i\lambda_k + m_u )
  {\prod_\ell} ( i\lambda_\ell + m_d ) \bigg\rangle^{\!\!\text{YM}}_{\Q}\,, 
  \label{eq:Z-nu}
\ea
where the average is taken with respect to the pure Yang-Mills action, and 
$|\Q|$ exact zero modes are implicitly included in the product. 
Equating the microscopic partition function \eqref{eq:Z-nu} to that of the effective theory
\eqref{eq:Znu} and taking their derivatives with respect to 
$m_u$ and/or $m_d$, one finds various nontrivial formulas for correlation functions 
of the Dirac eigenvalues. 
The simplest one is given by 
\ba
    \frac{\der}{\der m_u} \log Z_\Q & = \akakko{ \sum_{n} \frac{1}{i\lambda_n+m_u} }_\Q 
    \\
    & = \int_{-\infty}^{\infty}\dd \lambda~\frac{\akakko{\rho^A(\lambda)}_{\Q}}{i\lambda + m_u}
    \\
    & = 2V_4f_2 m_u + 2V_4f_A m_d 
    \frac{I_{\Q}'(2V_4f_Am_um_d)}{I_\Q(2V_4f_Am_um_d)} +\calO(m^3) \,,
    \label{eq:rho_re}
\ea
with $I_\Q'(x)\equiv \dd I_\Q(x)/\dd x$. Now $\akakko{\dots}_\Q$ stands for 
the average with full $N_f=2$ QCD measure. In the thermodynamic limit, the number of 
Dirac eigenvalues scales linearly with $V_4$, 
so let us define the one-point function (or the \emph{macroscopic} spectral density) by
\ba
  R_1(\lambda) & = \lim_{V_4\to \infty}\frac{1}{V_4}\akakko{\rho^A(\lambda)}_{\Q}\,,
\ea 
in terms of which \eqref{eq:rho_re} reads as
\ba
	  \int_{-\infty}^{\infty} 
	  \dd\lambda~\frac{R_1(\lambda)}{i\lambda + m_u} 
	  = 2 f_2 m_u + 2 f_A m_d + \calO(m^3) \,. 
	  \label{eq:rho_rel}
\ea
Note that $R_1(\lambda)$ depends on $m_u$ and $m_d$ implicitly through the 
QCD measure used for averaging. $R_1(\lambda)$ is expected to have no 
dependence on $\Q$ since topology is irrelevant once the thermodynamic limit 
is taken. Strictly speaking, we have to specify 
a UV cutoff scheme in order to make \eqref{eq:rho_rel} fully meaningful.   
Equation \eqref{eq:rho_rel} will be used later in Sec.~\ref{sec:comment}. 

There exists another relation that directly relates $R_1(\lambda)$ to the $\U(1)_A$ anomaly 
\cite{Chandrasekharan:1995gt,Chandrasekharan:1998yx}. 
Setting $m_u=m_d\equiv m$ and evaluating the chiral susceptibilities \eqref{eq:chis} 
in the basis of Dirac eigenstates, one can straightforwardly obtain 
\ba
    \chi_\pi = 4\int_{0}^{\infty} \frac{\dd \lambda}{\lambda^2+m^2}R_1(\lambda)
    \quad 
    \text{and}
    \quad 
    \chi_\delta = 4\int_{0}^{\infty}\dd \lambda~\frac{\lambda^2-m^2}{(\lambda^2+m^2)^2}R_1(\lambda) \,.
\ea
By subtraction we arrive at \cite{Chandrasekharan:1995gt,Chandrasekharan:1998yx}
\ba
    \chi_\pi - \chi_\delta & =  8 
    \int_{0}^{\infty}\dd \lambda~\frac{m^2}{(\lambda^2+m^2)^2}R_1(\lambda) \,. 
    \label{eq:chidiffe}
\ea
Combining this formula with \eqref{eq:c-dif}, we find 
\ba
    \int_{0}^{\infty}\dd \lambda~\frac{m^2}{(\lambda^2+m^2)^2}R_1(\lambda) 
    = f_A + \calO(m^2) \,.
    \label{eq:Rchichi}
\ea
It is clear from this relation that small Dirac eigenvalues are necessary 
for $f_A$ to be nonzero.%
\footnote{Large eigenvalues with density $R_1(\lambda)\sim \lambda^3$ 
only affect the $\calO(m^2)$ part; $f_A$ receives no contribution.} 
Indeed $f_A=0$ follows immediately if $R_1(\lambda)$ 
has a spectral gap near zero at $T>T_c$\,. 
Equations \eqref{eq:rho_rel}, \eqref{eq:chidiffe} and \eqref{eq:Rchichi} highlight 
essential properties of $R_1(\lambda)$. 

Unfortunately the form of $R_1(\lambda)$ in the near-zero region cannot be deduced 
uniquely from \eqref{eq:rho_rel} and \eqref{eq:Rchichi} alone. Actually there are 
infinitely many functions that satisfy \eqref{eq:Rchichi}; e.g.,
\ba
	R_1(\lambda) ~~ \sim ~~
	\frac{f_A m^{2n+1-k}\lambda^k}{(\lambda^2+m^2)^{n}}  
	\quad \text{and} \quad f_A m^2\delta(\lambda) \,, 
	\label{eq:R1examples}
\ea
where $n\geq 1$ and $0\leq k\leq 2n+1$ are arbitrary integers. 
Recently, the Dirac spectrum in two-flavor QCD at $T\gtrsim T_c$ 
has been studied intensively in lattice QCD simulations 
\cite{Bazavov:2012qja,Cossu:2013uua,Buchoff:2013nra,Dick:2015twa} 
(see also \cite{Bernard:1996iz,Kogut:1998rh,Chandrasekharan:1998yx,
Edwards:1999zm,Farchioni:1999ws,
Damgaard:2000cx,Gattringer:2001ia} for early works). 
The three possibilities $R_1(\lambda)\sim m$, $\lambda$, and $m^2\delta(\lambda)$ 
were examined in detail in 
\cite{Bazavov:2012qja,Buchoff:2013nra}, whereas the Breit-Wigner form 
$R_1(\lambda)\sim \rho_0 A/(\lambda^2+A^2)$ was nicely 
fitted to the lattice data in \cite{Dick:2015twa,Sharma:2015wua} 
(but see \cite{Cossu:2014aua,Tomiya:2014mma,Cossu:2015kfa,Cossu:2015lnb} 
for detailed investigations of lattice artifacts stemming from partial reweighting).  
The $\delta$ form is motivated 
by the dilute instanton gas picture \cite{Gross:1980br} in QCD at $T\to\infty$, but 
the exact $\delta$ form is unlikely to emerge at $T\gtrsim T_c$ due to the overlap 
of neighboring instantons and anti-instantons.  
For the moment, contrasting results from different simulations 
do not allow us to draw a definitive conclusion on the form of $R_1(\lambda)$.

\subsection{Spectral sum rules II:~microscopic limit}

There is yet another way to take the thermodynamic limit in \eqref{eq:rho_re}:  
if we let all of $\lambda$, $m_u$ and $m_d$ scale as $1/\sqrt{V_4 f_A}$ 
in the $V_4\to \infty$ limit, the dependence on the topology does persist. To see 
this, let us define a rescaled dimensionless spectral density 
\ba
    \rho_\Q(\zeta; \mu_u, \mu_d) 
    & \equiv \lim_{V_4\to\infty} \frac{1}{\sqrt{2 V_4 f_A}} 
    \akakko{
        \rho^A\mkakko{\frac{\zeta}{\sqrt{2 V_4 f_A}}}
        \bigg|_{m_u=\frac{\mu_u^{\mathstrut}}{\sqrt{2V_4f_A}},~
        m_d=\frac{\mu_d}{\sqrt{2V_4f_A}}}
    }_{\!\!\!\Q}\,, 
    \label{eq:rho_mic}
\ea 
which is analogous to the microscopic spectral density 
in the $\epsilon$-regime \cite{Leutwyler:1992yt,Verbaarschot:2000dy} but note that the relevant 
scale of eigenvalues here is $1/\sqrt{V_4}$ rather than $1/V_4$.%
\footnote{Intriguingly, a similar unusual scaling $\sim 1/\sqrt{V_4}$ also 
appears in color-superconducting phases of QCD at high density 
\cite{Yamamoto:2009ey,Kanazawa:PhD}, 
in the superfluid phase of two-color QCD \cite{Kanazawa:2009ks,Kanazawa:2009en} 
and in an exotic phase proposed by Stern 
\cite{Stern:1998dy,DescotesGenon:1999zj,Kanazawa:2015kca}.}  
Then \eqref{eq:rho_re} becomes 
\ba
    \int_{-\infty}^{\infty} \dd\zeta~
    \frac{1}{i\zeta + \mu_u}\rho_\Q(\zeta;\mu_u,\mu_d) 
    = \frac{f_2}{f_A} \mu_u + \mu_d \frac{I_{\Q}'(\mu_u\mu_d)}{I_\Q(\mu_u\mu_d)}\,. 
    \label{eq:micros}
\ea 
Notice that all $\calO(m^3)$ corrections in \eqref{eq:rho_re} drop out in this limit; 
hence \eqref{eq:micros} is \emph{exact}. This, of course, comes with a caveat that 
such a limit is meaningful only when the assumption $f_A\ne 0$ is correct. 
(See Appendix \ref{app:scaling} for another microscopic scaling.) 
It is straightforward to extend the rescaling \eqref{eq:rho_mic} to higher-order 
spectral correlation functions. 

We conjecture that spectral fluctuations on the scale $1/\sqrt{V_4}$ should be 
universal, i.e., determined solely by global symmetries and independent of 
the detailed form of QCD interactions in the ultraviolet. 
Although such a new ``microscopic limit'' prompts us to construct a 
random matrix theory that describes the Dirac spectrum in this regime, we have not been 
successful yet. The difficulty in finding a proper random matrix theory may have something to do 
with the fact that \emph{no} global symmetry is spontaneously broken at $T>T_c$\,, 
unlike in the QCD vacuum \cite{Leutwyler:1992yt,Verbaarschot:2000dy} 
and high-density QCD \cite{Yamamoto:2009ey,Kanazawa:2009ks,Kanazawa:2009en}.

We can derive infinitely many spectral sum rules rigorously, 
by expanding the following expression in powers of quark masses,  
\ba
  \frac{(m_um_d)^{-\Q}Z_\Q}{\displaystyle \lim_{m\to 0}\big[(m_um_d)^{-\Q}Z_\Q\big]} & = 
  \Q !\frac{I_\Q(2V_4f_Am_um_d)}{(V_4f_Am_um_d)^\Q}
  \ee^{V_4 f_2 (m_u^2+m_d^2)}
  \nonumber 
  \\
  & = \left\langle  
  {\prod_k}' \left(1+\frac{m_u^2}{\lambda_k^2}\right)
  {\prod_\ell}' \left(1+\frac{m_d^2}{\lambda_\ell^2}\right) \right\rangle_{\Q}\,,
  \label{eq:ewrfds}
\ea
where $\Q\geq 0$ is assumed and 
the product ${\prod}'$ runs only over $\lambda_n>0$. The average above is taken for the 
massless two-flavor QCD measure. 
Note that both sides of \eqref{eq:ewrfds} are normalized to unity in the chiral limit. 
Two examples of the sum rules read
\ba
  \left\langle{\sum_k}' \frac{1}{\lambda_k^2} \right\rangle_{\Q} \!=V_4f_2 
  \qquad \text{and} \qquad 
  \left\langle{\sum_k}' \frac{1}{\lambda_k^4} \right\rangle_{\Q} \!=
  \frac{(V_4f_A)^2}{1+\Q} \,.  
  \label{eq:sr}
\ea
${\sum}'$ above denotes a sum over $\lambda_n>0$.  
Note that the first sum rule receives UV-divergent contributions from large eigenvalues 
with density $\propto \lambda^3$, implying that 
$f_2$ is a regularization-scheme-dependent quantity.  By contrast, 
the second sum rule is dominated by contributions from $\calO(1/\sqrt{V_4})$ 
eigenvalues. UV eigenvalues only give $\calO(V_4)$ correction to the RHS, which 
is negligibly small as compared to the $\calO(V_4^2)$ term in the $V_4\to \infty$ limit.  
Indeed one can show that $f_A$ is free from UV divergences (see Appendix B of 
\cite{DescotesGenon:1999zj}).  
The suppression of the second sum rule for large $\Q$ is due to the 
repulsion of Dirac eigenvalues from the origin by $\Q$ exact zero modes. 

In terms of the ``microscopic'' spectral density, the sum rules \eqref{eq:sr} read
\begin{subequations}
\ba
    \int_{0}^{\infty}\dd \zeta~\frac{\rho_\Q(\zeta;0,0)}{\zeta^2} 
    & = \frac{f_2}{2f_A}\,, 
    \\
    \int_{0}^{\infty}\dd \zeta~\frac{\rho_\Q(\zeta;0,0)}{\zeta^4} 
    & = \frac{1}{4(1+\Q)}\,.
\ea
\end{subequations}
To obtain the universal function $\rho_\Q(\zeta;0,0)$ analytically is an important open problem 
that deserves further investigation.

\subsection{New Banks--Casher-type relation}
\label{sec:Banks--Casher}

If the axial anomaly is present at high temperature, it will be manifested not only in 
the Dirac eigenvalue density but also in the $n$-point spectral correlation 
functions for $n \geq 2$. To examine this possibility, let us introduce the 
\emph{connected two-point correlation function} 
(see e.g., \cite{Toublan:2000dn})%
\footnote{In the rest of this subsection we will ignore topological zero modes altogether. 
As argued in Sec.~\ref{sc:CSanom}, this is justifiable in the macroscopic limit 
with a positive path-integral measure.}   
\ba
	\label{eq:RC}
    R_{\rm C}(\lambda,\lambda') \equiv \lim_{V_4\to\infty}\frac{1}{V_4}
    \bigg[
        \akakko{\rho^A(\lambda)\rho^A(\lambda')} - \akakko{\rho^A(\lambda)}\akakko{\rho^A(\lambda')}
    \bigg] \,. 
\ea
$R_{\rm C}$ depends on $m_{u,d}$ implicitly through the averaging weight.  Note that 
$R_{\rm C}$ satisfies the constraint 
\ba
    \int_{-\infty}^{\infty}\dd \lambda~R_{\rm C}(\lambda,\lambda') = 
    \int_{-\infty}^{\infty}\dd \lambda'~R_{\rm C}(\lambda,\lambda') = 0 \,. 
    \label{eq:RCcondit}
\ea
If eigenvalues are entirely uncorrelated, they obey the Poisson statistics. 
In this case, the two-point function is related to the one-point function 
(cf. \cite[(5.4)]{Guhr:1996wn} and \cite[(3.33)]{Guhr:1997ve}) as 
\ba
    \akakko{\rho^A(\lambda)\rho^A(\lambda')}_{\rm Po}
    = 
    \ckakko{
      \delta(\lambda-\lambda') + \delta(\lambda+\lambda') 
    } \akakko{\rho^A(\lambda)} + \left( 1 - \frac{1}{N} \right) \akakko{\rho^A(\lambda)} 
    \akakko{\rho^A(\lambda')} , \!\!\!
    \label{eq:pois}
\ea
where $N$ denotes the number of chiral pairs of Dirac eigenvalues 
$\{\pm i \lambda_n\}$ (hence the total number of eigenvalues is $2N$). 
The $\delta$-functions in first term represent a trivial self-correlation. 
In Appendix \ref{ap:pois} we outline the derivation of \eqref{eq:pois} for completeness.   
Then we obtain for the uncorrelated Dirac spectra 
\ba
	R_{\rm C}^{\rm Po}(\lambda,\lambda') =
	\ckakko{
		\delta(\lambda-\lambda') + \delta(\lambda+\lambda') 
	} R_1(\lambda) - \frac{V_4}{N} R_1(\lambda)R_1(\lambda') \,. 
    \label{eq:poiss}
\ea
As $N$ grows linearly with $V_4$, $V_4/N$ has a well-defined thermodynamic limit.  
One can check that \eqref{eq:poiss} satisfies \eqref{eq:RCcondit}. 

Nontrivial two-level correlations among eigenvalues can be characterized by the deviation of $R_{\rm C}$ 
from the Poisson case. Let us define the \emph{two-point cluster function} $T_2(\lambda,\lambda')$ by
\ba
    T_2(\lambda,\lambda') \equiv
    R_{\rm C}^{\rm Po}(\lambda,\lambda') 
    - R_{\rm C}(\lambda,\lambda') \,.
    \label{eq:T2def}
\ea
We observe that $T_2(\lambda,\lambda')=T_2(\lambda',\lambda)$ and 
\ba
	  T_2(\lambda,\lambda')=T_2(-\lambda,  
	  \lambda')=T_2(\lambda,-\lambda') 
	  = T_2(-\lambda,-\lambda') 
	  \label{eq:rho_c_sym}
\ea
owing to chiral symmetry. For the Poisson distribution, 
$T_2$ vanishes identically by definition. 

To see how $R_{\rm C}$ and $T_2$ are related to the axial anomaly, we 
note that \eqref{eq:ch1} and \eqref{eq:ch3} together with $\chi_\sigma=\chi_\pi$ imply
\ba
    \chi_\pi-\chi_\delta & = 4 \chi_{\rm disc}  = 8 f_A + \calO(m^2) \,,
    \label{eq:xdisc}
\ea
where $\chi_{\rm disc}$ is the \emph{disconnected scalar susceptibility} defined by 
\ba
	  \chi_{\rm disc} & \equiv 
	  \lim_{V_4\to\infty}\frac{1}{V_4}
	  \frac{\der^2}{\der m_u \der m_d}\log Z
	  \nonumber 
	  \\
	  & = \lim_{V_4\to\infty}\frac{1}{V_4} 
	  \Bigg(
	  \bigg\langle\sum_{i}\frac{1}{i\lambda_i+m_u}\sum_{j}\frac{1}{i\lambda_j+m_d}
	  \bigg\rangle 
	  - \bigg\langle\sum_{k}\frac{1}{i\lambda_k+m_u}\bigg\rangle 
	  \bigg\langle\sum_{\ell}\frac{1}{i\lambda_\ell+m_d}\bigg\rangle 
	  \Bigg) 
	  \nonumber 
	  \\
	  & = \int_{-\infty}^\infty \dd\lambda 
	  \int_{-\infty}^\infty \dd\lambda' 
	  \frac{R_{\rm C}(\lambda,\lambda')}{(i\lambda+m_u)(i\lambda'+m_d)}\,.
	  \label{eq:intrho}
\ea
In the limit $m_u=m_d=m$, we substitute \eqref{eq:T2def} into \eqref{eq:intrho}, 
obtaining 
\ba
	\chi_{\rm disc} = &  
	\int_{-\infty}^\infty \dd\lambda 
	\kkakko{
	    \frac{R_1(\lambda)}{(i\lambda+m)^2} + \frac{R_1(\lambda)}{\lambda^2+m^2}
        } - \frac{V_4}{N} \mkakko{
            \int_{-\infty}^\infty \dd\lambda~\frac{R_1(\lambda)}{i\lambda+m}
        }^2 
        \notag
        \\
        & - \int_{-\infty}^\infty \dd\lambda 
	  \int_{-\infty}^\infty \dd\lambda' 
	  \frac{T_2(\lambda,\lambda')}{(i\lambda+m)(i\lambda'+m)}
        \\
        = & 
        \int_{0}^\infty \dd\lambda~\frac{4m^2}{(\lambda^2+m^2)^2} R_1(\lambda) 
        - \int_{-\infty}^\infty \dd\lambda 
	  \int_{-\infty}^\infty \dd\lambda' 
	  \frac{T_2(\lambda,\lambda')}{(i\lambda+m)(i\lambda'+m)} 
        + \calO(m^2)
        \\
        = & ~
        4 f_A - \int_{-\infty}^\infty \dd\lambda 
	  \int_{-\infty}^\infty \dd\lambda' 
	  \frac{T_2(\lambda,\lambda')}{(i\lambda+m)(i\lambda'+m)} 
        + \calO(m^2) \,,
\ea
where \eqref{eq:rho_rel} and \eqref{eq:Rchichi} have been used. 
Recalling \eqref{eq:xdisc} we obtain
\ba
      \int_{-\infty}^\infty \dd\lambda 
      \int_{-\infty}^\infty \dd\lambda' 
      \frac{T_2(\lambda,\lambda')}{(i\lambda+m)(i\lambda'+m)} 
      = 2 f_A + \calO(m^2) \,.
      \label{eq:T2}
\ea 

Finally, taking the chiral limit in (\ref{eq:T2}), 
we arrive at a new Banks--Casher-type relation
\ba
    T_2(0,0)  = \frac{2}{\pi^2}f_A  
    \label{eq:rhoc_BC}
\ea
for massless two-flavor QCD at $T>T_c$\,.  
To the best of our knowledge, \eqref{eq:rhoc_BC} is a new result. 
It reveals that the $\U(1)_A$ anomaly is encoded, 
not only in the spectral density as in \eqref{eq:Rchichi}, but also in the nontrivial two-level 
correlations among near-zero eigenvalues.  
If the near-zero Dirac eigenvalues are entirely uncorrelated, 
then $T_2$ vanishes and leads to $f_A=0$, suggesting 
effective restoration of the $\U(1)_A$ symmetry.  
Whether an analogue of \eqref{eq:rhoc_BC} can be derived 
in $N_f>2$ QCD at $T>T_c$ is an interesting open problem. 

Some remarks are in order. 
In taking the chiral limit we have replaced $1/[(i\lambda+m)(i\lambda'+m)]$ with 
$\pi^2 \delta(\lambda)\delta(\lambda')$. Strictly speaking, in doing so we have tacitly 
assumed that the typical scale over which $T_2(\lambda,\lambda')$ 
varies is much larger than $m$. Whether this is true or not in actual QCD is a dynamical 
problem and must be checked separately.%
\footnote{This smoothness condition is necessary to derive the original 
Banks--Casher relation, too \cite{Leutwyler:1992yt}.} 
We also remark that \eqref{eq:rhoc_BC} cannot be extended to $T<T_c$ 
because of the infrared-singular behavior 
$\displaystyle T_2(\lambda,\lambda')\sim 
- \frac{\Sigma^2}{32\pi^4F^4}\log|\lambda-\lambda'|$ \vspace{2pt} 
for $|\lambda - \lambda'|\ll \lambda$ 
\cite{Toublan:2000dn}.

One might suspect that the correlation in the Dirac spectra revealed by \eqref{eq:rhoc_BC} 
is at variance with the quasi-instanton picture proposed in 
\cite{Kanazawa:2014cua}  
where a Poisson distribution of topological objects 
(i.e., dressed instantons called \emph{quasi-instantons}) 
was argued at all $T>T_c$\,.%
\footnote{Topological objects similar to our quasi-instantons have 
been advocated for color-superconducting phases of QCD at high density \cite{Son:2001jm, Yamamoto:2008zw}. 
While quasi-instantons in hot QCD do not interact with each other 
\cite{Kanazawa:2014cua}, 
those in dense QCD weakly interact via exchange of (pseudo) 
Nambu-Goldstone modes \cite{Son:2001jm, Yamamoto:2008zw}.} 
In the limit $T \rightarrow \infty$, where the interaction is weak, 
the quasi-instanton gas is expected to reduce to the conventional 
dilute bare instanton gas \cite{Gross:1980br}. 
Let us try to explain how they can be consistent with each other. 
The point is that the Poisson distribution of topological zero modes (quasi-instantons) does 
\emph{not} necessarily mean the Poisson distribution of Dirac eigenvalues. 

In the quasi-instanton picture, independently distributed 
topological charges are expected to generate small Dirac eigenvalues 
that can be described, to a good approximation, by a spectral density with a $\delta$-peak at the origin.  
Let us discuss how to deal with this case explicitly within the present spectral analysis. 
The spectral density over a gauge field $A_\mu$ now assumes a form
\ba
    \rho^A(\lambda) = c^A \delta(\lambda) + \tilde\rho^A(\lambda)\,,
    \label{eq:rhonew}
\ea 
where $\tilde\rho^A(\lambda)$ is the density of eigenvalues 
away from zero and $c^A\geq 0$ is an integer which is equal to 
the total number of topological objects, $N = N_++N_-$.%
\footnote{Precisely speaking, the $c_A$ modes consist of $|\Q|=|N_+-N_-|$ exact zero modes 
and $\mathrm{Min}(N_+,N_-)$ chiral pairs of near-zero modes.} 
If we assume that the density of nonzero eigenvalues $\tilde\rho^A(\lambda)$ 
is so small that the anomalous contribution $f_A$ in \eqref{eq:Rchichi} 
solely originates from the $\delta$ peak at the origin, then it follows that  
\ba
    \akakko{c^A} & = 2 V_4 f_A m_u m_d\,. 
    \label{eq:cAeq}
\ea
(In deriving this we have used $\int_0^\infty \dd x\ \delta(x)=1/2$.) 
It is pleasing to see that \eqref{eq:cAeq} agrees with the quasi-instanton density derived in \cite[Sec.~V]{Kanazawa:2014cua}. 

Next, by plugging \eqref{eq:rhonew} into \eqref{eq:RC} we find 
\ba
    R_{\rm C}(\lambda,\lambda') = \ &  
    \frac{1}{V_4}\left[\left\langle(c^A)^2\right\rangle-\left\langle{c^A}\right\rangle^2 \right]
    \delta(\lambda)\delta(\lambda') + \tilde R_{\rm C}(\lambda,\lambda')
    \notag
    \\
    & + \frac{1}{V_4} \delta(\lambda) \left[ \left\langle c^A \tilde \rho^A(\lambda')\right\rangle - 
    \left\langle{c^A}\right\rangle \left\langle \tilde \rho^A(\lambda') \right\rangle \right]
    \notag
    \\
    & + \frac{1}{V_4} \delta(\lambda') \left[ \left\langle c^A \tilde \rho^A(\lambda)\right\rangle - 
    \left\langle{c^A}\right\rangle \left\langle \tilde \rho^A(\lambda) \right\rangle \right] \,,
    \label{eq:rccc}
\ea
where $\tilde R_{\rm C}$ is the two-point connected function for eigenvalues away from zero. 
The last two lines will vanish if there is no correlation between zero and nonzero modes, 
which we assume. Then we substitute \eqref{eq:rccc} into \eqref{eq:intrho} to obtain  
\ba
    \chi_{\rm disc} & = 
    \frac{1}{V_4 m_um_d}\left[\left\langle(c^A)^2\right\rangle-\left\langle{c^A}\right\rangle^2 \right]
    + \int_{-\infty}^\infty \dd\lambda 
	  \int_{-\infty}^\infty \dd\lambda' 
	  \frac{\tilde R_{\rm C}(\lambda,\lambda')}{(i\lambda+m_u)(i\lambda'+m_d)}\,.  
\ea
As we have been assuming that the density of nonzero modes is sufficiently low, it 
follows that the second term can be neglected in the chiral limit compared to the first term. 
Then \eqref{eq:cAeq} and $\chi_{\rm disc}=2f_A$ (cf.~\eqref{eq:xdisc}) 
imply
\ba
    \akakko{(c^A)^2}-\akakko{c^A}^2 = \akakko{c^A} = 
    2 V_4 f_A m_u m_d \,. 
\ea
This coincidence between the average and the variance of $c_A$ indicates 
that $c^A$ is Poisson distributed. This is indeed 
what the quasi-instanton picture in \cite{Kanazawa:2014cua} suggests. 

We mention that the Poisson statistics of topological objects was 
indeed observed in recent lattice data at $T=1.5T_c$ \cite{Dick:2015twa}. 
However, in the real world, quasi-instantons will not be strictly 
noninteracting (due to the $\calO(m^4)$ term in the free energy) 
and the $\delta$-peak 
of the spectral density may not be sufficiently narrow to rigorously justify the 
above treatment. Also, the correlations between zero modes and nonzero 
modes will not be negligible in general. With these caveats in mind, 
we still believe that the quasi-instanton picture in 
\cite{Kanazawa:2014cua} and the exact Banks--Casher-type relation \eqref{eq:rhoc_BC} can be a useful 
starting point for a fuller analytical and numerical investigation of the Dirac spectrum 
in QCD at high temperature in future.

\section{Comment on the Aoki--Fukaya--Taniguchi theorem}
\label{sec:comment}

Contrary to our assumption that $f_A \neq 0$ for $T>T_c$\,, Aoki \emph{at al.} 
\cite{Aoki:2012yj} claim that, under certain assumptions, the violation of the $\U(1)_A$ symmetry is invisible 
in correlation functions of scalar and pseudoscalar 
quark bilinears for $T>T_c$ in two-flavor QCD.  (This claim 
does not generalize to the vector--axial-vector sector, as we discussed in Sec.~\ref{sec:review}.)   
There are \emph{two key assumptions} in their analysis:%
\footnote{%
Aoki \emph{et al.} used overlap fermions on the lattice 
to regularize UV divergences. This is not crucial in the following discussion, however.}
\begin{enumerate}
  \item 
  The Dirac spectral density can be expanded in Taylor series 
  \beq
  	\label{rho_exp}
  	\displaystyle    
  	R_1(\lambda)=\sum_{n=0}^{\infty} \langle \rho^A_n \rangle_m\frac{\lambda^n}{n!} 
  \eeq
  near the origin, 
  with a radius of convergence that does not vanish in the chiral limit.  
  In particular, there is no $\delta(\lambda)$ term. The notation $\langle \rho^A_n \rangle_m$ 
  makes it clear that these coefficients are dependent on the quark mass $m$. 
  \item 
  The expectation value $\langle O(A) \rangle_m$ of any $m$-independent observable $O(A)$ is an analytic 
  function of $m^2$ at $T>T_c$\,. (Here $O(A)$ must be a functional of the gauge field only. It does not 
  include fermionic observables, such as the chiral condensate. The quarks must be integrated out 
  before this assumption is applied.)
\end{enumerate}
It should be noted that none of the examples in \eqref{eq:R1examples} 
satisfies the first assumption.

Precisely speaking, Ref.~\cite{Aoki:2012yj} assumes that the spectral density 
for a given gauge field $A_\mu$ can be expanded in Taylor series, while the above 
assumption 1 is only concerned with the spectral density averaged over all gauge fields.%
\footnote{We thank Sinya Aoki for clarifying this point to us.}

While their original proof \cite{Aoki:2012yj} is rather involved, we now show that 
a much simpler proof of $f_A=0$ for $N_f=2$ 
based on our analysis in the former sections 
is possible. Namely, one can easily prove the following theorem:
\paragraph{\underline{Theorem.}}
Under the two assumptions above, $f_A=0$. 

\paragraph{\underline{Proof.}}
From (\ref{rho_exp}) and the Banks--Casher relation \cite{Banks:1979yr}, we have  
\beq
	\langle\bar\psi\psi\rangle\propto\lim_{m\to 0}\langle\rho^A_0\rangle_m \,.
\eeq
For $T>T_c$ where $\langle\bar\psi\psi\rangle = 0$, it must be that
\beq
	\lim_{m\to 0}\langle\rho^A_0\rangle_m=0\,. \label{eq:rhozero}
\eeq
Since $\langle\rho^A_0\rangle_m$ is an analytic function of $m^2$ according to 
the second assumption, \eqref{eq:rhozero} means that 
$\langle\rho^A_0\rangle_m=\calO(m^2)$; in particular
\ba
	\lim_{m_u\to 0} \langle\rho^A_0\rangle_m & = \calO(m_d^2) \,. 
	\label{eq:rho00}
\ea
On the other hand, it follows from \eqref{eq:rho_rel} that
\ba
  \int^{\Lambda}_{0} \dd\lambda~\frac{2m_u}
  {\lambda^2+m_u^2} R_1(\lambda)
  = 2 f_2 m_u + 2 f_A m_d + \calO(m^3)  \,,
  \label{eq:matching}
\ea
where a UV cutoff $\Lambda$ was inserted. 
Note that, to derive this expression, we have only used (i) analyticity of 
the free energy and (ii) irrelevance of exact zero modes in the thermodynamic limit 
(as explained in Sec.~\ref{sec:topology}). If the limit $m_u\to 0$ 
is taken with $m_d$ fixed, the RHS of \eqref{eq:matching} converges to 
$2f_A m_d+\calO(m_d^3)$\,.%
\footnote{The $\calO(m_d^3)$ contribution comes from 
higher-order $\U(1)_A$-violating terms (e.g.,  
$\tr(MM^\dagger)\det M+\text{c.c.}$) in the free energy.} 

Next, we introduce an arbitrary scale $\epsilon>0$ which is 
smaller than the radius of convergence 
of \eqref{rho_exp} in the chiral limit. Then we split the LHS of 
\eqref{eq:matching} as
\ba
	  \int^{\Lambda}_{0} \dd\lambda~\frac{2m_u}
	  {\lambda^2+m_u^2}R_1(\lambda) 
	  & = \int^{\epsilon}_{0} \dd\lambda~\frac{2m_u}
	  {\lambda^2+m_u^2}R_1(\lambda)
	  + \int_{\epsilon}^{\Lambda} \dd\lambda~\frac{2m_u}
	  {\lambda^2+m_u^2}R_1(\lambda)
	  \\
	  & \leq \int^{\epsilon}_{0} \dd\lambda~\frac{2m_u}
	  {\lambda^2+m_u^2}
	  \left[\sum_{n=0}^{\infty} \langle\rho^A_n\rangle_m\frac{\lambda^n}{n!} 
	  \right] 
	  + \frac{2m_u}{\epsilon^2}
	  \int_{\epsilon}^{\Lambda} \dd\lambda~R_1(\lambda)\,,
	  \label{eq:rhosplit}
\ea
where \eqref{rho_exp} was used. The second term 
in \eqref{eq:rhosplit} is obviously $\calO(m_u)$, whereas 
the first term is more nontrivial. To check its behavior near the chiral limit, 
we use
\begin{subequations}
  \ba
	  \int_0^\epsilon \dd \lambda~\frac{\lambda}{\lambda^2+m_u^2} 
	  & = - \log \frac{|m_u|}{\epsilon} + \calO(m_u^2)\,, 
	  \label{eq:mlead}
	  \\
	  \int_0^\epsilon \dd \lambda~\frac{\lambda^2}{\lambda^2+m_u^2} 
	  & = \epsilon - \frac{\pi}{2}|m_u| + \calO(m_u^2)\,,
	  \\
	  \int_0^\epsilon \dd \lambda~\frac{\lambda^3}{\lambda^2+m_u^2} 
	  & = \frac{\epsilon^2}{2}  + m_u^2 \log \frac{|m_u|}{\epsilon} + \calO(m_u^4)\,,
	  \\
	  \int_0^\epsilon \dd \lambda~\frac{\lambda^n}{\lambda^2+m_u^2} 
	  & = \frac{\epsilon^{n-1} }{n-1}+\calO(m_u^2)
	  \qquad \text{for}~n\geq 4\,.
  \ea
\end{subequations}
As the leading term in the limit  $m_u\to 0$ comes from \eqref{eq:mlead}, we deduce
\ba
  \int^{\epsilon}_{0} \dd\lambda~\frac{2m_u}
  {\lambda^2+m_u^2}
  \left[\sum_{n=0}^{\infty} \langle\rho^A_n\rangle_m\frac{\lambda^n}{n!} 
  \right] 
  & = 
  \int^{\epsilon}_{0} \dd\lambda~\frac{2m_u}
  {\lambda^2+m_u^2}\langle\rho^A_0 \rangle_m +
  \calO(m_u \log m_u)
  \\
  & = \big[ \pi+\calO(m_u) \big] \langle\rho^A_0 \rangle_m 
  +  \calO(m_u \log m_u) \,. 
  \label{eq:rhocafsd}
\ea
Plugging this into \eqref{eq:rhosplit}, we observe that 
\ba
	\int^{\Lambda}_{0} \dd\lambda~\frac{2m_u}
	{\lambda^2+m_u^2}R_1(\lambda) \leq 
	\pi \langle\rho^A_0 \rangle_m  + \calO(m_u \log m_u) \,. 
\ea 
Recalling \eqref{eq:rho00}, it is now clear that
\ba
	\lim_{m_u\to 0}\int^{\Lambda}_{0} \dd\lambda~\frac{2m_u}
	{\lambda^2+m_u^2}R_1(\lambda) = \calO(m_d^2)\,. 
\ea
This is to be compared with \eqref{eq:matching}, which tells that 
the leading term in the limit $m_u\to 0$ is $2 f_A m_d$.  
Thus $f_A=0$ is concluded. This completes the proof. 
\\\par 
This short proof is made possible by treating $m_u$ and $m_d$ as two independent variables, 
unlike the original one \cite{Aoki:2012yj}, where only the case of degenerate masses was considered. 
Of course, whether the two assumptions are correct or not in QCD is highly nontrivial. 
If $f_A$ is non-vanishing in QCD, which has been assumed in the former sections, 
then one has to relax at least one of the two conditions above. Considering that 
recent lattice simulations \cite{Bazavov:2012qja,Buchoff:2013nra,Dick:2015twa} 
have demonstrated a singular peak structure in $R_1(\lambda)$ at small $\lambda$, 
it seems natural to abandon the Taylor expansion \eqref{rho_exp}. This issue deserves 
further investigation.

\section{Conclusions}
\label{sec:conclusion}
In this paper, we derived some rigorous results on the violation of the $\U(1)_A$ 
symmetry in two-flavor QCD at $T>T_c$\,, which is characterized by the difference of chiral susceptibilities, 
$\chi_{\pi}-\chi_{\delta}$ (see (\ref{eq:c-dif})) and is parametrized by 
$f_A$ (see (\ref{eq:Zex}) for the definition). 
We clarified how the different topological sectors conspire to violate 
the $\U(1)_A$ symmetry and how it varies with the spatial volume of the system. 
We demonstrated that any moment of the topological charge at $T>T_c$ can be obtained, 
once just a single parameter $f_A$ is fixed.
We also derived new spectral sum rules and a Banks--Casher-type relation that  
relate the anomaly strength $f_A$ to statistical correlations in Dirac spectra.
As a by-product of the sum rules, we found a simple proof of the Aoki--Fukaya--Taniguchi 
``theorem'' on the effective restoration of the $\U(1)_A$ symmetry \cite{Aoki:2012yj}. 
Since nontrivial assumptions are required to prove this theorem, we cannot conclude 
$\U(1)_A$ restoration in QCD yet. However, our simplified proof would hopefully serve to 
understand the importance of these assumptions more clearly. 

All of our new exact relations can, in principle, be tested on the lattice. In particular, the 
relation (\ref{eq:chichi00}) can be used to extract the value of $f_A$ at $T>T_c$ 
even in a small volume with fixed topology ($\Q=0$). 
Finally, we note that determination of $f_A$ should also be important from a phenomenological 
point of view, as it is related, through (\ref{chi}), to the temperature-dependent mass 
of the QCD axion---an input for the evolution of the axion density, which might account 
for the dark matter density of the universe; 
see, e.g., \cite{Berkowitz:2015aua, Kitano:2015fla} for recent works.

 \acknowledgments 
 The authors thank S.~Aoki, M.~Buchoff, N.~Christ, G.~Cossu, 
 K.~Fukaya, M.~Laine, S.~Sharma and Y.~Taniguchi  
 for useful discussions and correspondences. 
  This work was supported, in part, by the RIKEN iTHES Project, 
  JSPS KAKENHI Grants Number 26887032, and
  MEXT-Supported Program for the Strategic Research Foundation
  at Private Universities, ``Topological Science'' (Grant Number S1511006).

\appendix 

\section{Tensor decomposition of anomalous correlators at finite temperature}
\label{app:tensor}

Consider the anomalous three-point function in QCD at finite temperature 
$T$ in momentum space,
\beq
  \label{eq:correlation}
  T^{\sigma \rho \mu}(q, p, T)
  \equiv
  \!\int\! \dd^4 x\, \dd^4 y\ 
  \ee^{iqx+ipy} \big\langle j^{\sigma}(x) j^{\rho}(y) 
  j^{A\mu}(0) \big\rangle,
\eeq
where $j^{\mu}=\bar \psi \gamma^{\mu} \psi$ is the vector current and
$j^{A\mu} = \bar \psi \gamma^{\mu} \gamma^5 \psi$ is the axial current.
The color and flavor degrees of freedom are suppressed for simplicity. 
We take the rest frame of the medium as $\eta^{\mu}=(1,\bm{0})$.

Itoyama and Mueller wrote down 30 tensor invariants composed of 
$q^{\mu}$, $p^{\nu}$, $\eta^{\kappa}$, or $\epsilon^{\alpha \beta \gamma \delta}$ 
for $T^{\sigma \rho \mu}$ \cite{Itoyama:1982up}. However, their expression is not 
complete; one can write down 30 more tensor invariants. 
We find that the most general decomposition for $T^{\sigma \rho \mu}$ 
is given by
\beq
  T^{\sigma \rho \mu}&=&
  A_1 q_{\tau} \epsilon^{\tau \sigma \rho \mu}
  + A_2 p_{\tau} \epsilon^{\tau \sigma \rho \mu}
  + A_3 \eta_{\tau} \epsilon^{\tau \sigma \rho \mu}
  \nonumber 
  \\
  & &
  + A_4 q^{\sigma}q_{\alpha}p_{\beta}
  \epsilon^{\alpha \beta \rho \mu}
  + A_5 p^{\sigma}q_{\alpha}p_{\beta} 
  \epsilon^{\alpha \beta \rho \mu}
  + A_6 \eta^{\sigma}q_{\alpha}p_{\beta} 
  \epsilon^{\alpha \beta \rho \mu}
  \nonumber 
  \\
  & &
  + A_7 q^{\rho} q_{\alpha}p_{\beta} 
  \epsilon^{\alpha \beta \sigma \mu}
  + A_8 p^{\rho} q_{\alpha}p_{\beta} 
  \epsilon^{\alpha \beta \sigma \mu}
  + A_9 \eta^{\rho} q_{\alpha}p_{\beta} 
  \epsilon^{\alpha \beta \sigma \mu}
  \nonumber 
  \\
  & &
  + A_{10} q^{\sigma}q_{\alpha}\eta_{\beta} 
  \epsilon^{\alpha \beta \rho \mu}
  + A_{11} p^{\sigma}q_{\alpha}\eta_{\beta} 
  \epsilon^{\alpha \beta \rho \mu}
  + A_{12} \eta^{\sigma}q_{\alpha}\eta_{\beta} 
  \epsilon^{\alpha \beta \rho \mu}
  \nonumber 
  \\
  & &
  + A_{13} q^{\rho}q_{\alpha}\eta_{\beta} 
  \epsilon^{\alpha \beta \sigma \mu}
  + A_{14} p^{\rho}q_{\alpha}\eta_{\beta} 
  \epsilon^{\alpha \beta \sigma \mu}
  + A_{15} \eta^{\rho}q_{\alpha}\eta_{\beta} 
  \epsilon^{\alpha \beta \sigma \mu}
  \nonumber 
  \\
  & &
  + A_{16} q^{\sigma}q_{\alpha}\eta_{\beta} \epsilon^{\alpha \beta \rho \mu}
  + A_{17} p^{\sigma}q_{\alpha}\eta_{\beta} \epsilon^{\alpha \beta \rho \mu}
  + A_{18} \eta^{\sigma}q_{\alpha}\eta_{\beta} \epsilon^{\alpha \beta \rho \mu}
  \nonumber 
  \\
  & &
  + A_{19} q^{\rho}p_{\alpha}\eta_{\beta} \epsilon^{\alpha \beta \sigma \mu}
  + A_{20} p^{\rho}p_{\alpha}\eta_{\beta} \epsilon^{\alpha \beta \sigma \mu}
  + A_{21} \eta^{\rho}p_{\alpha}\eta_{\beta} \epsilon^{\alpha \beta \sigma \mu}
  \nonumber 
  \\
  & &
  +\left[A_{22} q^{\sigma}q^{\rho} + A_{23} p^{\sigma}q^{\rho}
  + A_{24} \eta^{\sigma}q^{\rho} + A_{25} q^{\sigma}p^{\rho}
  +A_{26} p^{\sigma}p^{\rho} \right.
  \nonumber 
  \\
  & &
  \left.
  \ \ \ +A_{27}\eta^{\sigma}p^{\rho} + A_{28}   
  q^{\sigma}\eta^{\rho}
  +A_{29}p^{\sigma}\eta^{\rho} + A_{30} \eta^{\sigma}  
  \eta^{\rho}
  +A_{31}g^{\sigma \rho} \right]
  q_{\alpha} p_{\beta} \eta_{\gamma} 
  \epsilon^{\alpha \beta \gamma \mu}
  \nonumber 
  \\
  & &
  + A_{32}q^{\mu}q_{\alpha}p_{\beta}
  \epsilon^{\alpha \beta \sigma \rho}
  +A_{33}p^{\mu}q_{\alpha}p_{\beta}\epsilon^{\alpha \beta \sigma \rho}
  +A_{34}\eta^{\mu}q_{\alpha}p_{\beta}\epsilon^{\alpha \beta \sigma \rho}
  \nonumber 
  \\
  & &
  +A_{35}q^{\mu}q_{\alpha}\eta_{\beta}\epsilon^{\alpha \beta \sigma \rho}
  +A_{36}p^{\mu}q_{\alpha}\eta_{\beta}\epsilon^{\alpha \beta \sigma \rho}
  +A_{37}\eta_{\mu}q_{\alpha}\eta_{\beta}\epsilon^{\alpha \beta \sigma \rho}
  \nonumber 
  \\
  & &
  +A_{38}q^{\mu}p_{\alpha}\eta_{\beta}\epsilon^{\alpha \beta \sigma \rho}
  +A_{39}p^{\mu}p_{\alpha}\eta_{\beta}\epsilon^{\alpha \beta \sigma \rho}
  +A_{40}\eta^{\mu}p_{\alpha}\eta_{\beta}\epsilon^{\alpha \beta  \sigma \rho}
  \nonumber 
  \\
  & &
  +\left[A_{41} q^{\mu}q^{\rho} + A_{42} p^{\mu}q^{\rho}
  + A_{43} \eta^{\mu}q^{\rho} + A_{44} q^{\mu}p^{\rho}
  +A_{45} p^{\mu}p^{\rho} \right.
  \nonumber 
  \\
  & &
  \left.
  \ \ \ +A_{46}\eta^{\mu}p^{\rho} + A_{47} q^{\mu}\eta^{\rho}
  +A_{48}p^{\mu}\eta^{\rho} + A_{49} \eta^{\mu} \eta^{\rho}
  +A_{50}g^{\mu \rho} \right]
  q_{\alpha} p_{\beta} \eta_{\gamma} \epsilon^{\alpha \beta \gamma \sigma}
  \nonumber 
  \\
  & &
  +\left[A_{51} q^{\sigma}q^{\mu} + A_{52} p^{\sigma}q^{\mu}
  + A_{53} \eta^{\sigma}q^{\mu} + A_{54} q^{\sigma}p^{\mu}
  +A_{55} p^{\sigma}p^{\mu} \right.
  \nonumber 
  \\
  & &
  \left.
  \ \ \ +A_{56}\eta^{\sigma}p^{\mu} + A_{57}   
  q^{\sigma}\eta^{\mu}
  + A_{58}p^{\sigma}\eta^{\mu} 
  + A_{59} \eta^{\sigma} \eta^{\mu}
  + A_{60}g^{\sigma \mu} \right]
  q_{\alpha} p_{\beta} \eta_{\gamma} 
  \epsilon^{\alpha \beta \gamma \rho}
\eeq
with coefficients 
$A_i \equiv A_{i}({\bm q}^2,\,{\bm p}^2, \,{\bm k}^2, \,q \cdot \eta,\, p \cdot \eta, \,T)$.
The terms $A_{31, \cdots, 60}$ are new, which have been missed in 
\cite{Itoyama:1982up} but are generally allowed by symmetries.
It turns out that $A_{34}$, $A_{37}$, $A_{40}$, $A_{43}$, $A_{46}$, 
$A_{48}$, $A_{49}$, $A_{53}$, $A_{57}$, $A_{58}$, and $A_{59}$ are not 
independent of the others appearing above and can be omitted without loss of generality.%
\footnote{
For example, there exists an identity for the $A_{46}$ term, 
\beq
  \label{eq:identity}
  q_{\alpha} p_{\beta} p^{\rho} \eta_{\gamma} \eta^{\mu}
  \epsilon^{\alpha \beta \gamma \sigma}
  = - q_{\alpha} p_{\beta} p^{\rho} \epsilon^{\alpha \beta \sigma \mu}
  + (p \cdot \eta) q_{\alpha} p^{\rho} \eta_{\beta}
  \epsilon^{\alpha \beta \sigma \mu}
  + (q \cdot \eta) p_{\beta} p^{\rho} \eta_{\alpha}
  \epsilon^{\alpha \beta \sigma \mu}
  + q_{\alpha} p_{\beta} p^{\rho} \eta_{\gamma} \eta^{\sigma}
  \epsilon^{\alpha \beta \gamma \mu},  \nonumber  \\
\eeq 
with which the $A_{46}$ term can be expressed in terms of the $A_8$, $A_{14}$, $A_{20}$, and $A_{27}$ terms.} 
In particular, we need not consider tensor invariants 
whose axial-vector index $\mu$ is carried by $\eta^{\mu}$.
Therefore, there are 49 independent invariants in total.

\section{Another microscopic scaling}
\label{app:scaling}

In \eqref{eq:rho_mic}, all dimensionful quantities are rescaled by $\sqrt{2V_4f_A}$. 
On the other hand, it is also allowed (from a mathematical point of view) 
to use $\sqrt{2V_4f_2}$ for rescaling. Defining   
\ba
    \rho_\Q(\zeta; \mu_u, \mu_d) 
    & \equiv \lim_{V_4\to\infty} \frac{1}{\sqrt{2 V_4 f_2}} 
    \akakko{
        \rho^A\mkakko{\frac{\zeta}{\sqrt{2 V_4 f_2}}}
        \bigg|_{m_u=\frac{\mu_u}{\sqrt{2V_4f_2}},~
        m_d=\frac{\mu_d}{\sqrt{2V_4f_2}}}
    }_\Q\,, 
\ea 
we obtain from \eqref{eq:rho_re} a modified spectral relation
\ba
  \int_{-\infty}^{\infty} \dd\zeta~
  \frac{1}{i\zeta + \mu_u}\rho_\Q(\zeta;\mu_u,\mu_d) 
  = \mu_u + \frac{f_A}{f_2}\mu_d 
  \frac{I_{\Q}'\left(\frac{f_A}{f_2}\mu_u\mu_d\right)}
  {I_\Q\left(\frac{f_A}{f_2}\mu_u\mu_d\right)}\,. 
\ea
We suspect, however, that this scheme is precarious because $f_2$ is dominated by UV-divergent contributions from 
the perturbative Dirac spectra $R_1(\lambda)\sim \lambda^3$ which generally depends on the regularization scheme. 
In contrast, $f_A$ is free from UV divergences (see Appendix B in \cite{DescotesGenon:1999zj}). This leads us to consider 
the microscopic scaling by $f_A$ as the most natural one.

\section{Derivation of \eqref{eq:pois}}
\label{ap:pois}

The relation \eqref{eq:pois} for uncorrelated Dirac spectra can be easily shown as follows. 
With $2N$ Dirac eigenvalues $\{\pm i\lambda_n\}_{n=1}^N$ 
we have, from \eqref{eq:rhodef},
\ba
	\akakko{
		\rho^A(\lambda)\rho^A(\lambda')
	} & = \akakko{
		\sum_{k=1}^{N} \ckakko{
			\delta(\lambda-\lambda_k) + \delta(\lambda+\lambda_k)
		}
		\sum_{\ell=1}^{N} \ckakko{
			\delta(\lambda'-\lambda_\ell) + \delta(\lambda'+\lambda_\ell)
		}
	}
	\\
	& = \akakko{
		\sum_{k=1}^{N} \ckakko{
			\delta(\lambda-\lambda_k) + \delta(\lambda+\lambda_k)
		}\ckakko{
			\delta(\lambda'-\lambda_k) + \delta(\lambda'+\lambda_k)
		}
		}
	\notag
	\\
	& \quad ~ + \sum_{k=1}^{N} 
	\akakko{
			\ckakko{
				\delta(\lambda-\lambda_k) + \delta(\lambda+\lambda_k)
			}
			\sum_{\ell\ne k}^{} \ckakko{
				\delta(\lambda'-\lambda_\ell) + \delta(\lambda'+\lambda_\ell)
			}
	}
	\\
	& = \ckakko{\delta(\lambda-\lambda')+\delta(\lambda+\lambda')} 
	\akakko{\rho^A(\lambda)} 
	\notag
	\\
	& \quad~ + \sum_{k=1}^{N}
	\Big\langle 
		\delta(\lambda-\lambda_k) + \delta(\lambda+\lambda_k)
	\Big\rangle
	\akakko{
		\sum_{\ell\ne k}^{} \ckakko{
			\delta(\lambda'-\lambda_\ell) + \delta(\lambda'+\lambda_\ell)
		}
	}.
\ea
In the last step we factorized the average, which is justified by the 
absence of correlations among different eigenvalues.  
Then, using the trivial identity 
\ba
	\akakko{
		\sum_{\ell\ne k}^{} \ckakko{
			\delta(\lambda'-\lambda_\ell) + \delta(\lambda'+\lambda_\ell)
		}
	} = \frac{N-1}{N} \akakko{\rho^A(\lambda')} \,,
\ea
we arrive at the desired formula
\ba
	\akakko{
		\rho^A(\lambda)\rho^A(\lambda')
	} & = \ckakko{\delta(\lambda-\lambda')+\delta(\lambda+\lambda')} 
	\akakko{\rho^A(\lambda)} + \frac{N-1}{N}
	\akakko{\rho^A(\lambda)}\akakko{\rho^A(\lambda')}\,. 
\ea
One can see by integrating over $\lambda$ and $\lambda'$ that 
both sides are normalized to $4N^2$ correctly.

\bibliography{draft_anomaly_resub_v3.bbl}
\end{document}